\documentstyle[graphics,graphicx,]{l-aa}
\begin{document}

   \thesaurus{04(10.01.1; 10.03.1; 10.07.1; 10.07.3; 10.19.2) }

   \title{Age and metallicity gradients in the Galactic Bulge \thanks{B
ased on observations with the NASA/ESA Hubble Space Telescope,
      obtained at the Space Telescope Science Institute, which is
operated by
      the Association of Universities for Research in Astronomy,
Inc. under
      NASA contract No. NAS5-26555}}

   \subtitle{A differential study using HST/WFPC2 }

   \author{Sofia Feltzing \inst{1,2,3} \and Gerard Gilmore \inst{2}}

   \offprints{Sofia Feltzing}
  \institute{Royal Greenwich Observatory
              Madingley Road,
              Cambridge CB3 0EZ,
              U.K.
   \and Institute of Astronomy,Madingley Road,Cambridge CB3 0HA
   \and Present address: Lund Observatory, Box 43, S-221 00 Lund, Sweden
   }
 
%   \authorrunninghead{S. Feltzing \& \G. Gilmore}
%   \titlerunninghead{Age and metallicity gradients in the Galactic
%   Bulge}

   \date{Recieved 06-03-1998; accepted 22-12-1999}

   \maketitle

   \begin{abstract}

The Galactic Bulge has long been assumed to be a largely old stellar
population. However, some recent studies based on observations with
the HST WF/PC-1 and WFPC2 of stars in the Galactic Bulge have
concluded that the old population may not make up more then 30\% of
the total. Other studies using HST/WFPC2 differential studies of `Bulge'
globular clusters and field stars have found the bulge to be
comparable in age to the Galactic Halo. A complication in all these
studies is the presence of a substantial population of stars which
mimic a young bulge population, but which may be, and are often
assumed to be, foreground disk stars whose reddening and distance
distributions happen to mimic a young bulge turnoff. We show, using
number counts in HST/WFPC2 colour-magnitude diagrams of both field
stars in the Bulge and of two `bulge' and one `disk' globular cluster
(NGC6528, NGC6553, and NGC5927) that the stars interpreted as young in
fact are foreground disk stars. Thus, we confirm that the bulk of the
bulge field stars in Baade's Window are old. The existence of a young {\sl
metal-rich} population cannot, however, be ruled out from our data.

We also test for age and metallicity gradients in the Galactic Bulge
between the two low extinction windows Baade's window ($\ell$=$1\fdg1,
b$=$-4\fdg8$) and Sagittarius-I ($\ell$=$1\fdg3, b$=$-2\fdg7$). We use
the colour-magnitude diagram of a metal-rich globular cluster as an
empirical isochrone to derive a metallicity difference of $\la 0.2$
dex between Baade's window and SGR-I window. This corresponds to a
metallicity gradient of $\la 1.3$ dex/kpc, in agreement with recent
near-IR CMD studies. Such a steep gradient, if detected, would require
the existence of a short scale length inner component to the Bulge, most
likely that prominent in the near infra red, which perhaps forms a
separate entity superimposed on the larger, optical Bulge as observed
in Baade's window.


\end{abstract}
%
%________________________________________________________________

\section{Introduction}
The nature and origin(s) of Galactic Bulges are key aspects of any
galaxy formation model, and of the Hubble galaxy classification
sequence.  However, the present-day properties of bulges in
spiral galaxies are not well known (eg Silk \& Wyse 1993; Wyse,
et al. 1997).  Do bulges form late, early or continuously?
Are bulges related to halos? To disks? Are they single stellar
populations?  The existence of smooth and/or discontinuous gradients
in age and/or metallicity in the stellar population(s) in the Galactic
Bulge can help to discriminate between these different scenarios, and
is the topic of this paper.

Is there evidence for the widely repeated assumption that the Galactic
Bulge is old? Recently published colour magnitude
diagrams from HST/WFPC2 and HST/WFC1 of the Galactic Bulge and the
bulge globular clusters (Vallenari et al. 1996, Ortolani et al. 1995,
Holtzman et al 1993) while dominated by fairly old stars, show a
substantial population of stars above the dominant old turnoff.
Are these foreground disk stars, or is there a minority very young
bulge population? Since even a minority young bulge population is of
interest, we examine here the limits on young stars in the bulge
windows.

One common approach to determine the properties of the Bulge is to
study the so called bulge globular clusters (see e.g. Zinn 1996,
Ortolani et al. 1995, Minniti 1996). While primarily
motivated by observational convenience, the rationale behind this
approach is the assumption that these clusters may be valid tracers of
the stellar population(s) of the Bulge (see however Zinn 1996 and
Harris 1998) Formation scenarios relevant to this approach include the
possibility that the Bulge has been assembled from numerous such
clusters and these are the last surviving (Gnedin \& Ostriker 1997),
or that the clusters formed a system associated with the Bulge rather
than with the rest of the spheroidal component(s) of the Galaxy. Thus
the idea is that we may be able to infer the age and/or metallicity of
the Bulge stellar population either directly from studies of these
clusters, or through differential studies of the clusters and field
stars, under the assumption of similar metallicity. Since there is no
ab initio understanding of the formation of either galactic bulges or
globular clusters, and the age range of the globular cluster system
remains a topic of active debate, such analyses merit close
scrutiny. The most recent and extensive such analysis is that of
Ortolani et al. (1995, 1996) who observed two such globular clusters,
NGC6553 and NGC6528, with HST/WFPC2, and deduced that the Bulge has
the same age as the Halo.

Analysis of suitably-chosen globular clusters introduces
several possible complications.  The first is the major problem of
defining a proper population of `bulge' globular clusters. Some of
the clusters used, e.g.  Ter7, have recently been shown to be
associated with the satellite dwarf galaxy Sgr dSph, rather than with
the Galaxy itself.  They may therefore not be representative of the
Galactic Bulge.  The method of comparing ridge-lines of globular
clusters to infer relative ages requires the clusters in question to have
similar metallicities, and relative chemical abundances of the
alpha-elements, to avoid an age-metallicity degeneracy
 (Stetson et al. 1996, VandenBerg et al. 1990, 1996). 
New results by Cohen et al. (1999) show that NGC6553 may be as
much as $\sim 0.5$ dex more metal-rich than 47 Tuc, illustrating the
potentially large effects of metallicity range. The impressive recent
study of Rosenberg et al. (1999), indicating a  dispersion in ages for
the intermediate metallicity globular clusters, and  a large systematic age
difference between the metal-rich and metal-poor clusters illustrates
the complexity. 

Another potential
 uncertainty is that the metallicity distribution function of the
stars in the galactic Bulge is more similar to that in the
solar-neighbourhood (cf. Wyse and Gilmore 1995) than to that for
clusters within 5 degrees of the Galactic centre (Minniti 1996; Barbuy
et al. 1998). The bulge cluster distribution is both more narrow and
less metal-rich than the bulge field stars, complicating any direct
comparison.

Clearly, if one wishes to know the age of the bulge field stars, it is
desirable to observe the stars in the Bulge directly.  Direct studies
of the Bulge are difficult due to the severe crowding towards the
central regions of the Galaxy and the large, patchy, reddening along
the line of sight.  Several detailed studies of the outer Galactic
Bulge exist, providing kinematics and chemical abundance distribution
functions (Ibata and Gilmore 1995a, 1995b; Minniti et al. 1995; see
also Wyse et al. 1997). For the inner Bulge several
analyses of the low reddening Baade's window are available
(e.g. Ortolani et al. 1995, Vallenari et al. 1996, Holtzman et
al. 1993, Terndrup 1988, Ng et al. 1996), with direct studies of the
inner bulge field stars in the near-IR recently also becoming
available (Frogel et al. 1999), and even mid-IR ISO photometry (Omont
et al. 1999, Glass et al. 1999). Additionally, many recent studies have
emphasized the high continuing rate of star formation in the inner
bulge/disk. Do these stars diffuse with time the few hundred parsecs
into the Sgr and Baade's windows?

The interpretation of extant data is unclear, with a variety of
contradictory results.  Vallenari et al. (1996) use a mixture of
WF/PC-1 and NTT data while Holtzman et al. (1993) rely exclusively
on (the same) WF/PC-1 data.  Both groups reached the conclusion that
the Bulge is dominated by a significant young stellar
populations. Ortolani et al. (1995) using similar NTT data for Baade's
Window found the Bulge to be as old as the Halo.

The confusion among the results from the space based observations
should be contrasted with ground based optical observations which find
little evidence for a substantial young stellar population(s) in the
Galactic Bulge (eg Terndrup 1988), albeit rather far from the centre.
Note however that these observations do not cover the main-sequence
turn-off and the results are based on the giant branch. The
main-sequence turn-off is more sensitive to detection of a
significantly younger stellar populations.  Further complication is
provided by studies of OH/IR stars, suspected to be of intermediate
age, which are common in the inner bulge, or disk (Sevenster et al. 1997).

The distribution function of chemical abundances is a key parameter
defining a stellar population. In the outer Galactic Bulge Ibata \&
Gilmore (1995b) derived the relevant distribution function from
spectroscopy of K giants. They found a mean abundance of $\sim -0.2$
dex, with a very wide dispersion. In Baade's window McWilliam \& Rich
(1994) and Sadler et al. (1996) provided similar results.  Sadler et
al. (1996) found for 400 K giants a mean abundance of [Fe/H] $=-0.11
\pm0.04$ dex, with more than half the sample in the range $-0.4 < {\rm
[Fe/H]} < +0.3 {\rm dex}$.  This is similar to the results from the
detailed spectroscopic analysis of McWilliam \& Rich (1994).  A
metallicity gradient has been suspected for fields outside Baade's
Window, Terndrup (1988) and Minniti et al. (1995).  The important
conclusion from the spectroscopic analyses is that the stellar
populations of the inner Galactic Bulge are complex, and that their
analysis requires careful consideration of projected disk and other
populations.

In this paper we study colour-magnitude diagrams, derived from
archival HST/WFPC2 images, for 4 fields and two clusters towards the
Galactic Bulge and one ``disk'' cluster. We perform a purely
differential study of the properties of the Galactic Bulge
population(s), quantifying any systematic offsets and/or gradients in
age and/or metallicity in the field population(s).

The paper is organized as follows; in Sect. 2 we detail the
observations used and in Sect. 3 describe how we derive the
photometric magnitudes from the images. Sect. 4 discusses reddening
and distances for the individual fields and clusters, and the utility
of the cluster data. Sect. 5 discusses the age of the Bulge, while
Sect. 6 asks the question whether a metallicity gradient might be
present in the inner Bulge. Sect. 7 includes a summarizing discussion
which puts our results into the context of other studies. A brief
summary is found in Sect. 8.
%
%________________________________________________________________
\section{The data}

\begin{figure*}
\resizebox{12cm}{!}{\includegraphics[angle=-90]{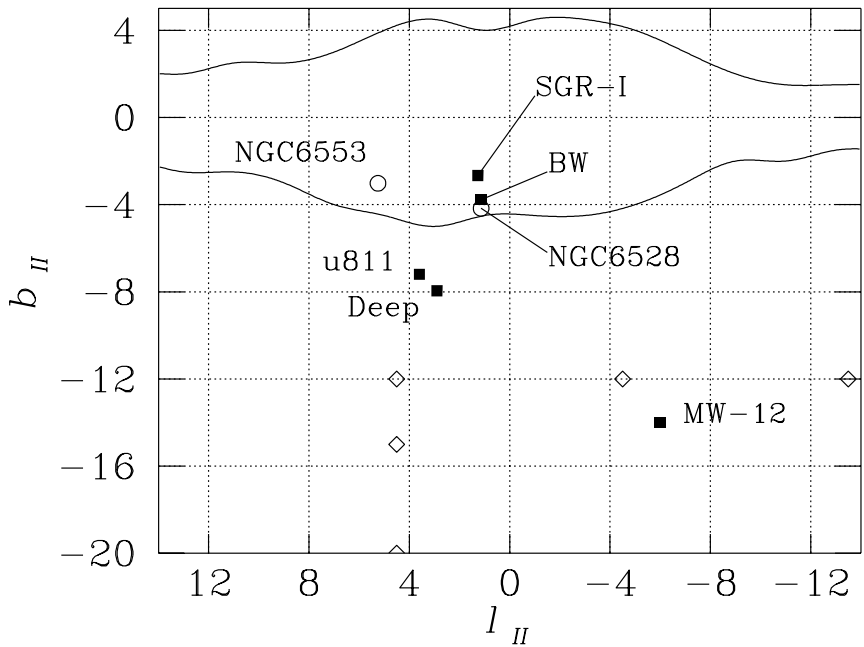}}
\hfill
\parbox[t]{55mm}{
\caption[]{Positions of fields and clusters.  Clusters are denoted by
open circles and fields by filled squares.  NGC5927 (with
$l=-34\degr$) is well outside this map.  We also show the positions of
the 5 inner-most fields studied by Ibata and Gilmore (1995a,b),
denoted by open diamonds. The Sgr dSph has its major axis along
$l=+5\degr$, and extends down to $b \la 4\degr$}
\label{field_positions}}
\end{figure*}

All observations analyzed here were obtained from the HST-archive.
The observations for the two clusters, NGC6528 and NGC6553, have
previously been reported and discussed in Ortolani et al.\,(1995) and
the observation of NGC5927 in Fullton et al.\,(1996).  However, to get
a set of data which is consistently treated we have derived our own
photometry from the original images.  The images are detailed in Table
\ref{fieldlist}, and their positions on the sky are shown in
Fig. \ref{field_positions}.  The results presented here are all, due
to the crowding in some of the fields, based exclusively on the PC1
images.

\subsection{Data reductions} 

All frames have been recalibrated through the STScI pipeline
calibration for HST at the Space Telescope - European Coordinating
Facility, using the most suitable flat field and bias frames
available.

\subsection{Combining images}

Because the WFPC2 images are under-sampled extra care has to be
exercised so that counts in the centre of stars are not lost when
images are combined in order to remove cosmic rays.  Sub-pixel shifts
between two images can cause the centre flux in one image (or in both)
to look like a cosmic ray when the images are compared in the cosmic
ray algorithm. Experiments on several of the image sets, using the
{\tt{stsdas.hst\_calib.wfpc.crrej}} task, showed that up to 10\% of
the counts may be lost unless a term linear in the counts (scalenoise)
is included in the modeling of $\sigma$ used in the rejection
algorithm (see the help file for {\sc{crrej}}).

\begin{table*}
\caption[]{Coordinates and passbands of observations for the fields
and clusters. For the fields we give the coordinates for the centre of
the PC1 and for the clusters the coordinates for the cluster
centres. the column headed ID give the HST archive identification
number of the original observing program in which the observations
were obtained. The last column give the total number of stars
simultaneously detected in F814 and F555W or F606W according to the
selection criteria discussed in Sect.\ref{sect:cmd}. If the data were
truncated the truncation magnitude is indicated in the last column.}
\begin{tabular}{lrrllllrllllllllll}
\hline\noalign{\smallskip}
Field &$l$ & $b$ & Passbands & Total exp.time& ID & Date of obs. &\multicolumn{2}{l}{ F814W+F555W/F606W} \\
\noalign{\smallskip}
\hline\noalign{\smallskip}
SGR-I   & 1\fdg27 &-2\fdg66 & F814W, F555W & 3000,3000 &5207 & 28/08/94 & 2940 \\
BW      & 1\fdg14 &-3\fdg77 & F814W, F555W & 2000,2000 & 5105&12/08/94& 2221& ($V_{\rm 555}<26$)\\
        &      &      & F814W, F555W & 80,80     & 6185& 2/09/95  & 1150 &($V_{\rm 555}<23$)\\
MW-12   & -6\fdg00   & -14\fdg00  & F814W, F555W & 1000,1200 &6614 & 9/10/96 & 63\\
u811    & 3\fdg60  & -7\fdg20 & F814W, F606W & 2600,2600 &5371 & 14,15/09/97& 907\\
Deep    & 2\fdg90  & -7\fdg95& F814W, F606W & 5800,2900&6254  &  1/05/96 & 463& ($V_{\rm 606}<27$)\\ 
NGC5927 & 326\fdg6 & 4\fdg86& F814W, F555W & 3200,2400 &5366 & 8/05/94  & 3289\\
        &  (-34\degr)&     & F814W, F555W & 210,150   &5366 &8/05/94   & 3587\\
NGC6528 & 1\fdg14 &-4\fdg17 & F814W, F555W & 200,100   &5436 & 27/02/94 & 1824\\
NGC6553 & 5\fdg25 &-3\fdg02 & F814W, F555W & 200,100   &5436 & 25/02/94 &2795 \\
\noalign{\smallskip} 
\hline
\end{tabular}
\label{fieldlist}
\end{table*}

%____
%________________________________________________________________

\section{Stellar Photometry:
Deriving colour-magnitude diagrams}
\label{sect:cmd}

To derive photometric magnitudes we have used the standard {\sc{iraf}}
\footnote{IRAF is distributed by National Optical Astronomy
Observatories, operated by the Association of Universities for
Research in Astronomy, Inc., under contract with the National Science
Foundation, USA.}  tasks in the {\sc{daophot}} package. To calibrate
our data we have used the procedures detailed in Holtzman et
al.\,(1995b), as well as empirical determinations of aperture
corrections.

We follow the general procedure of first detecting possible stars with
{\sc{daofind}}, derive initial photometric magnitudes from aperture
photometry using {\sc{phot}}, derive an analytical
point-spread-function ($psf$) for each image (assumed constant over
the entire chip) and finally perform $psf$-fitting on the list of
possible stars to determine stellar magnitudes using
{\sc{allstar}}. The initial aperture photometry is done in an aperture
of 2 pixels. We use the $psf$-fitted magnitudes in our analysis. The
raw $psf$-fitted magnitudes need to be corrected for a number of
effects which are peculiar to HST, and calibrated onto the HST
in-flight system.

For the observations at $(l,b)=(2\fdg9,-7\fdg95)$ only one observation
in the F606W passband was available. We identified the objects on the
combined F814W image and used that list to identify the stars on which
to perform measurements in the F606W image. The colour-magnitude
diagram was then searched for anomalous looking stars, which were
inspected by eye in both passbands and if deemed to be contaminated by
cosmic rays, excluded.

We first present the steps in our calibration and then discuss and
detail each step separately as some of the steps are non-trivial. Our
calibration contains the following steps;

\begin{enumerate}
\item apply empirical correction for the difference 
between aperture and $psf$-fitted magnitudes,
\item add 2 electrons to the flux in each pixel 
inside a radius of 5 pixels,
\item correct for the CTE effect,
\item correct for the geometric distortion,
\item normalize to WF3 and bay 4 (if applicable),
\item apply Holtzman's synthetic aperture corrections from 5 pixels out to
11 pixels (0$\farcs{5}$),
\item add synthetic zero points from Holtzman et al.
\end{enumerate}

\subsection{Aperture vs $psf$-photometry ($ap/psf$)}

\begin{figure}
\resizebox{\hsize}{!}{\includegraphics{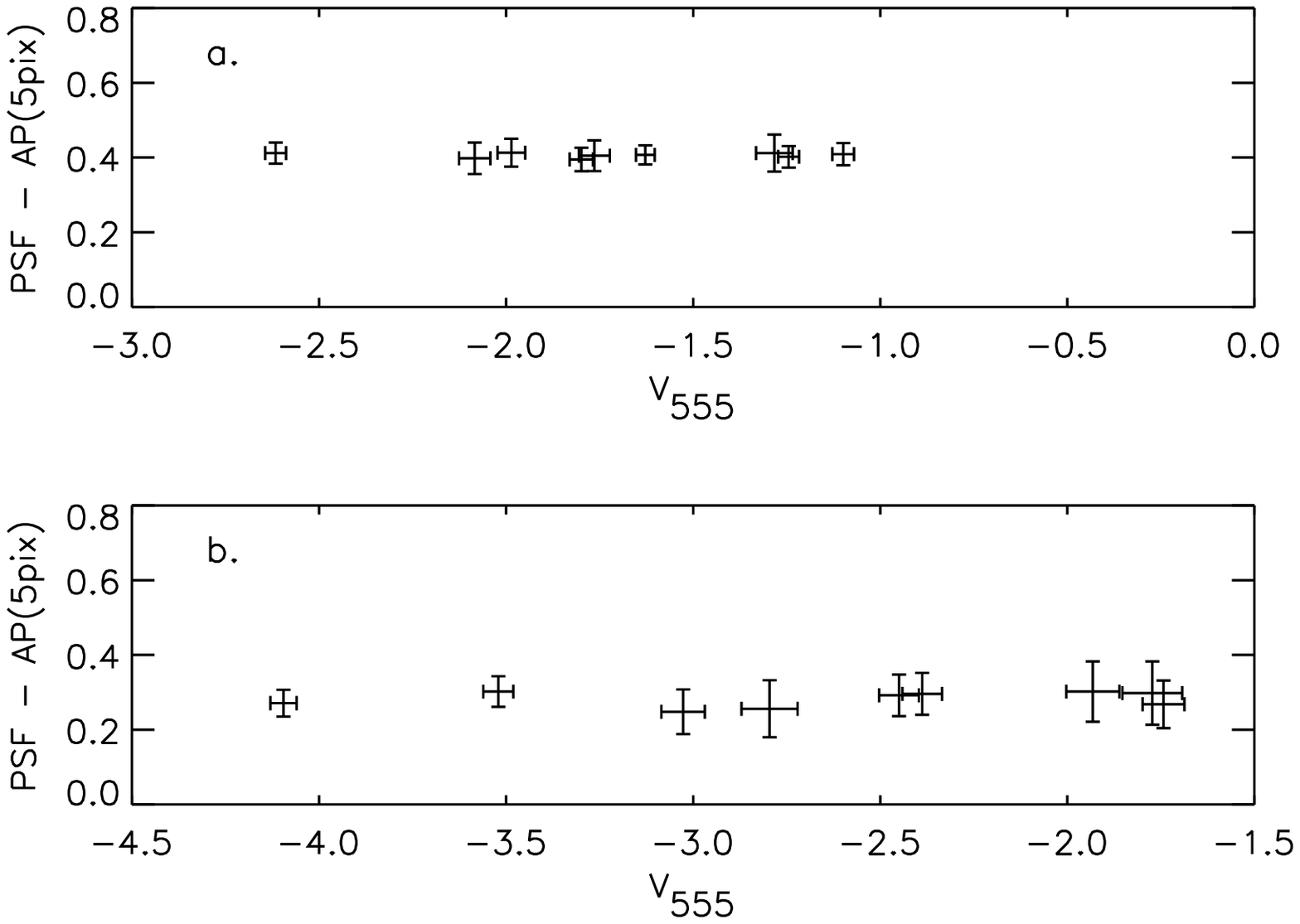}}
\caption[]{Difference between $psf$-fitted magnitudes and aperture
magnitudes measured inside an aperture with radius 5 pixels as a
function of the $psf$-fitted magnitudes. In panel a. we show the data
for Baade's window long exposure and in panel b. short exposure. The
error bars show the errors as given by {\sc {phot}} and {\sc
{allstar}}. Note the different ranges on the x-axes.}
\label{diff_mag}
\end{figure}

\begin{table}
\caption[]{Zero points, aperture corrections from Holtzman et al. (1995a) and
corrections for the zero points of the empirical $psf$s.}
\begin{tabular}{llllllllllllllllll}
\hline\noalign{\smallskip}
 Quantity & Field & F555W & F606W & F814W\\
\noalign{\smallskip}
\hline\noalign{\smallskip}
Zero point & & 21.723 & 22.084 & 20.844\\
Ap.corr. &  & \underline{0.89} & \underline{0.88} & \underline{0.87}\\ 
(Holtzman)& & {0.96}&     {0.955} &   {0.95  }    \\
\noalign{\smallskip}
\hline\noalign{\smallskip}
$Psf$-        & SGR-I   &0.220  & --   & 0.460 \\
 zero-point   &BW long  & 0.296 & --   & 0.406 \\
 corr.        &BW short &0.281  & --   & 0.470 \\
              &MW-12    & 0.342 & --   & 0.509 \\
              &u811     & --    &0.361 & 0.485 \\
	      & Deep    & --    &0.363 & 0.434 \\
	      &NGC5927l & 0.303 & --   & 0.493 \\
	      &NGC5927s & 0.240 & --   & 0.469 \\
              &NGC6528  &0.270  & --   & 0.444 \\
              &NGC6553  &0.460  & --   & 0.310 \\
\noalign{\smallskip}
\hline
\end{tabular}
\label{ap_corr}
\end{table}

There is usually a zero-point offset (as well as a spread) between
magnitudes derived from $psf$-fitting and from aperture measurements.
Since we construct our own $psf$ from the frame itself this offset is
due to the magnitude assigned to the $psf$ by the {\sc {psf}}
task. This magnitude is the magnitude of the first star used in
constructing the $psf$, thus it is somewhat arbitrary. This also means
that the offset between aperture photometry and $psf$-fitted
photometry may not always be the same or even have the same sign in
two frames.

For the stars that had been used to create the analytic $psf$ we
measured aperture magnitudes inside 5 pixels and subsequently
calculated the difference between these magnitudes and the magnitudes
derived from $psf$-fitting, Fig. \ref{diff_mag}. The scatter around
the mean-value for these corrections is $\leq 0.03$ magnitudes both in
F555W and F814W and for both long and short exposures.  The difference
was used as an empirical aperture correction. The results are listed
in Table \ref{ap_corr}. The $psf$-magnitudes were in this way
corrected out to $0\farcs5$.

\subsection{Long versus short exposures}

For the field in Baade's window a set of images with exposure times of
2$\times$1000s, 4$\times$200s and 2$\times$40s is available. This has
made it possible to investigate to what extent the magnitudes derived
from long and short exposures of the same stars (and in this case also
in a crowded and reddened field) give the same magnitudes. There is a
well-known ``feature'' of HST photometry that short- and long-exposure
magnitudes do not agree in zero-point. An addition of $2 e^-/pixel$ to
the flux measured in aperture photometry has been suggested as an
empirical way of correcting for such a discrepancy. We confirm that
this is a practical empirical formula. As we correct out to 5 pixels
radius for the discrepancy between aperture and $psf$-fitted
magnitudes we add 2 electrons to the flux in each pixel inside that
radius. \footnote{Since this article was first submitted there has
been significant development in the understanding of the long-vs-short
exposure problem.  Further discussion of this problem and its solution
may be found in the WFPC2 Instrument Science Reports 98-02 which can
be found at the following URL {\tt
http://www.stsci.edu/ftp/instrument\_news/WFPC2\\/wfpc2\_bib.html}}

\subsection{Aperture corrections, Charge 
Transfer Efficiency ($CTE$) and Effective pixel area ($EPA$)}

 Aperture corrections to $0\farcs5$ apertures, i.e. 11 pixels on the
PC1, from the 5 pixel apertures used to derive the difference between
$psf$ and aperture magnitudes are employed from Table 2 of Holtzman et
al. (1995a).

 The $CTE$ has the effect that stellar objects in a given row (more
charge transfer) appear fainter then they should have appeared had
they been at a lower row number. This is corrected for through a
simple formula suggested by Holtzman et al. (1995a,1995b) which means
that at the highest row the detected light is 4\% higher than at the
first row and this correction changes linearly along the column. For
observations taken before the cool down of the CCDs in WFPC2, 32 April
1994, the CTE is higher, $\sim 10 \%$.

 As discussed in Holtzman et al. (1995a) the WFPC2 cameras have
geometric distortions. We correct for these distortions using the
information in the map of effective pixel area given in Fig.16,
Holtzman et al.\,(1995a).

\subsection{Completeness}

\begin{figure}
\centerline{\resizebox{7cm}{!}{\includegraphics{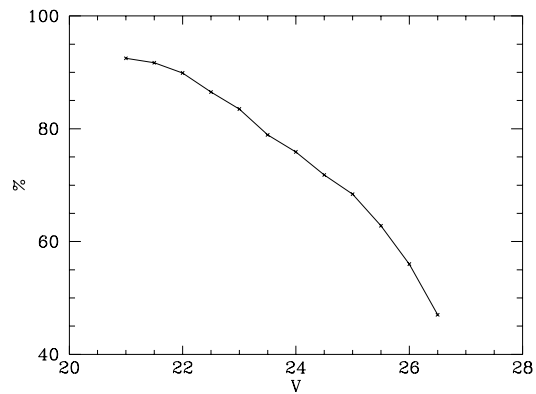}}}
\caption[]{Completeness for the deep exposure of Baade's window.
SGR-I shows the same level of completeness. This completeness is based
on the stars detected in both $V_{\rm 555}$ and $I_{\rm 814}$}
\label{compl}
\end{figure}

In order to determine the level of completeness, as a function of
derived apparent magnitude, we have performed the ``{\sc{addstar}}
experiment'', ie we add artificial stars at a given magnitude, thus
creating a new image.  The new frames were then put through the same
processes as described above and the list of synthetic stars created
by {\sc{addstar}} was cross-correlated with the final list of detected
stars to reveal how many of the synthetic stars had been
recovered. For this we required not only that the coordinates should
be the same but also that the new magnitude should be within 0.5
magnitudes of the magnitude assigned to the synthetic star by
{\sc{addstar}}, as illustrated in Fig. \ref{compl}.

Note that in this study we do not require a detailed knowledge of the
completeness function. This is because we are studying the turn off
regions and brighter in the colour-magnitude diagrams. Our only
requirement here is to show that completeness is not a significant
problem near the main-sequence turnoff. In fact, our photometry
extends several magnitudes fainter than necessary for this experiment.

\subsection{Selection of final stellar samples}

\begin{figure*}
\resizebox{\hsize}{!}{\includegraphics{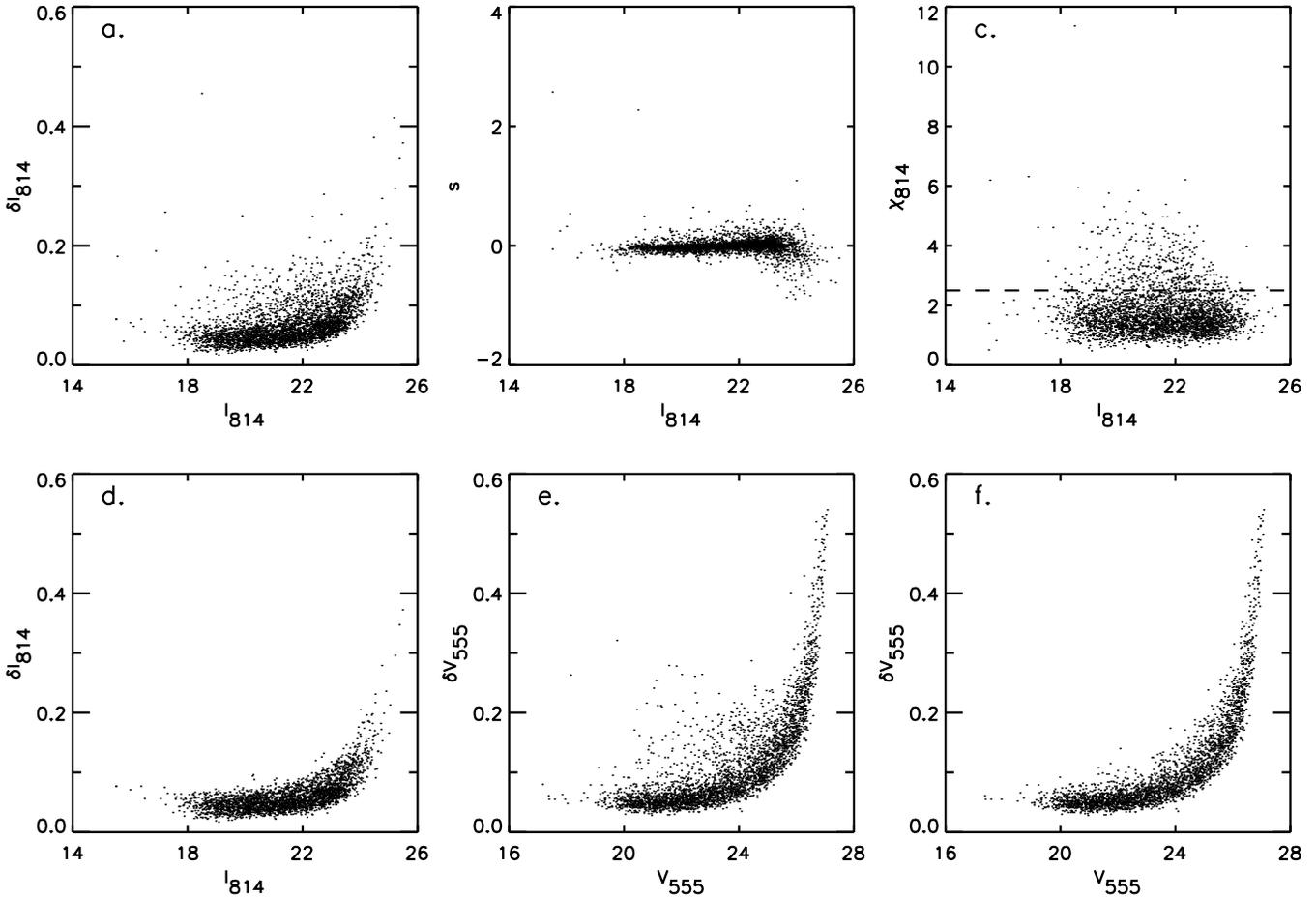}}
\caption[]{Diagnostics for the SGR-I field. Figures a., b., and
c. show the error, $\delta I_{\rm {814}}$, the sharpness and the
$\chi_{\rm {814}}$ as a function of the $ I_{\rm {814}}$ magnitude. In
d. we show the same as in a. but when the cut in $\chi_{\rm {814}}$ is
imposed (as indicated in c.). Finally in Figures e. and f.  $\delta
V_{\rm {555}}$ are shown as functions of $V_{\rm {555}}$
magnitudes. In e. without any cuts imposed and in f. when $\chi_{\rm
{555}}$ cut at $2.5$. Figure d. and f. thus show the distribution of
the errors in the final sample of stars in SGR-I colour-magnitude
diagram as displayed in Fig. \ref{cmd_deep}.}
\label{diagnostics}
\end{figure*}

All images were carefully inspected for saturated stars and spurious
detections around saturated stars (mottling) and along the diffraction
patterns. It was found that once the data for F814W and F555W were
matched very few saturated stars remained in the final sample and
virtually no detections on diffraction spikes and in the mottling
pattern remained.

The final selection of stellar samples was based on the diagnostic
diagrams produced from the $psf$-fitting photometry of which an
example is presented in Fig. \ref{diagnostics}. As can be seen in this
figure a cut at $\chi_{\rm {814}} < 2.5$ and $\chi_{\rm {555}} < 2.5$
cleans up the error vs. magnitude diagrams satisfactorily.

\begin{figure*}
\resizebox{\hsize}{!}{\includegraphics{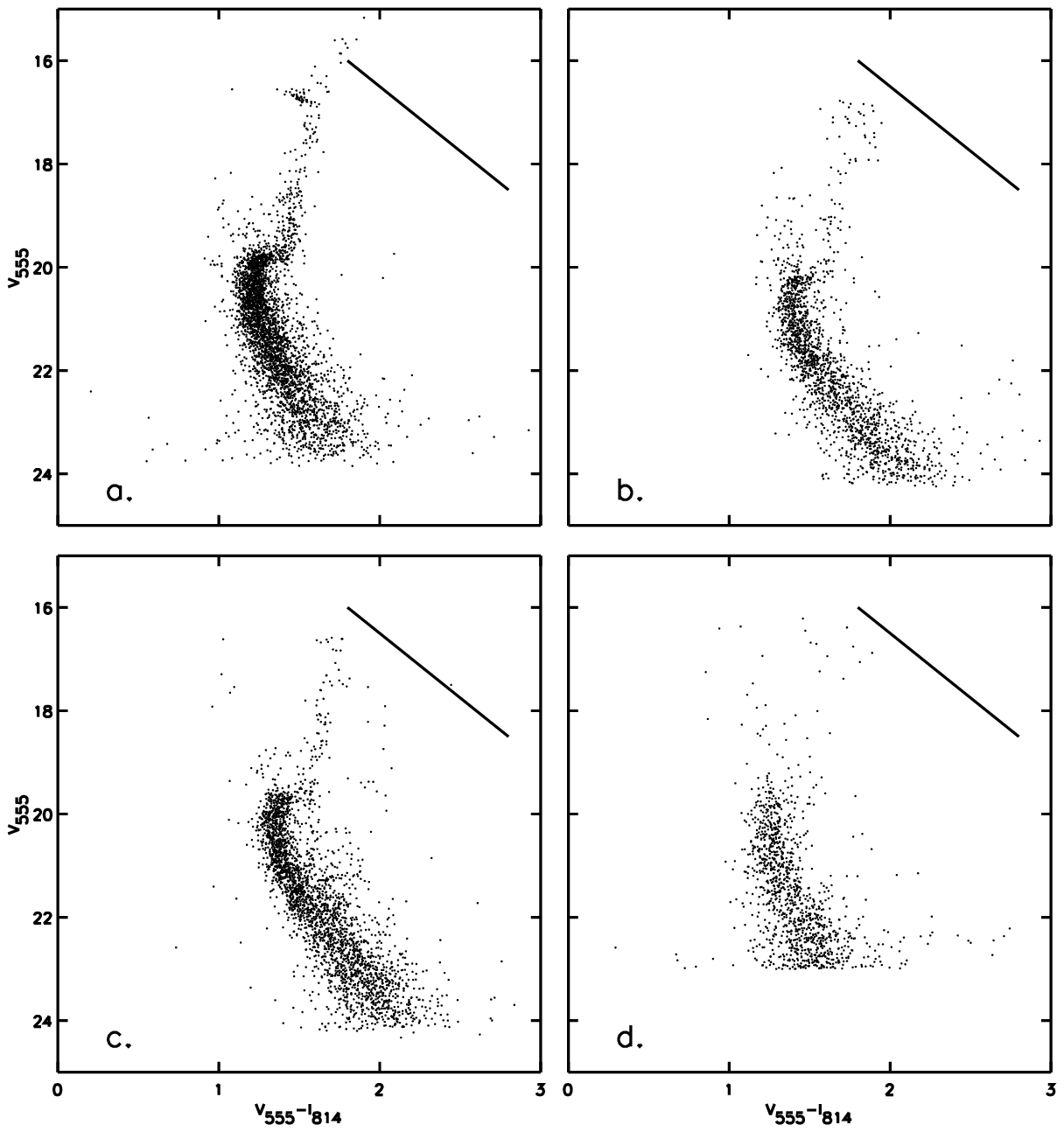}}
\caption[]{Colour magnitude diagrams for the cluster fields and the
short exposure of Baade's window.  a.  NGC5927 short exposure,
b. NGC6528 c. NGC6553, and d. BW short exposure.  The direction of the
reddening vector, as given in Holtzman et al. (1995b), is indicated by
a solid line in each diagram. The colour-magnitude diagram for Baade's
window was truncated at $V_{\rm 555} = 23$. }
\label{cmd_shallow}
\end{figure*}

\begin{figure*}
\resizebox{\hsize}{!}{\includegraphics{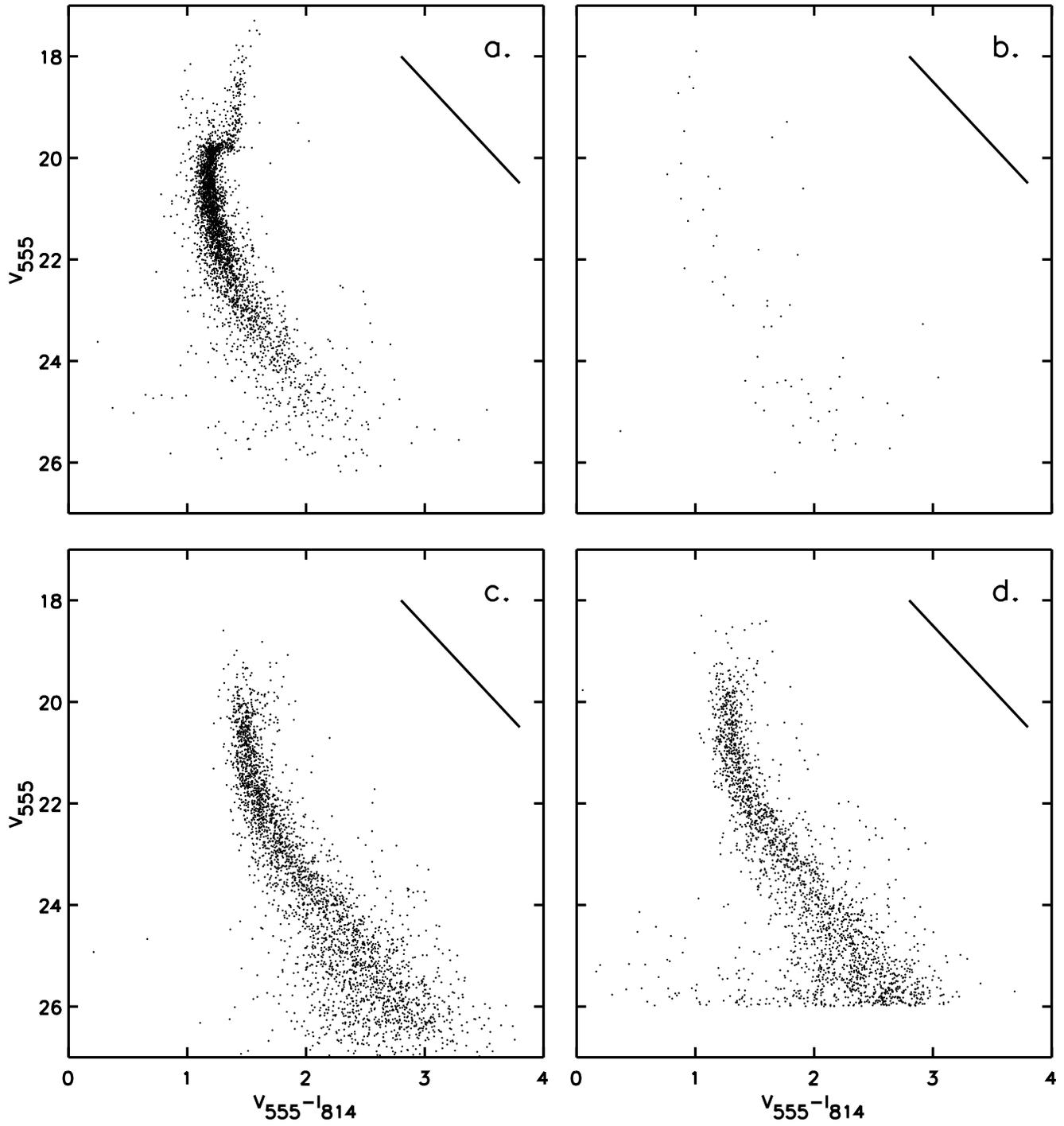}}
\caption[]{Colour magnitude diagrams for the two fields and NGC5927.
a.  NGC5927 long exposure, b. MW-12, c.  SGR-I, and d. BW long
exposure.  The direction of the reddening vector, as given in Holtzman
et al. (1995b), is indicated by a solid line in each diagram. }
\label{cmd_deep}
\end{figure*}

\begin{figure*}
\resizebox{\hsize}{!}{\includegraphics{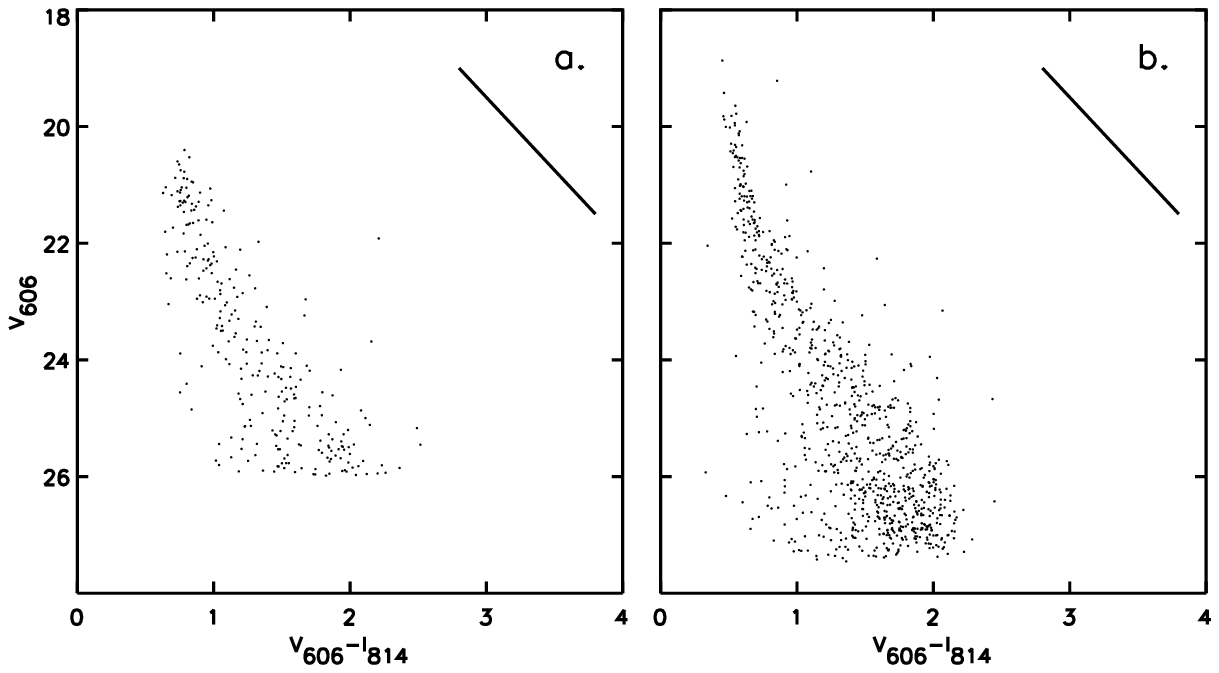}}
\caption[]{Colour magnitude diagrams for the two fields observed in
F606W and F814W. a. is the field at (l,b)=(2.9,-7.95) and b. at
(3.6,-7.2). The colour-magnitude diagram for the field at (2.9,-7.95)
has been truncated at $V_{\rm 606}=26$.  The direction of the
reddening vector, as given in Holtzman et al. (1995b), is indicated by
a solid line in each diagram. }
\label{cmd_606}
\end{figure*}

\section{Reddening, distances and metallicities.}
 
As we are using archive data we do not have control over which
passbands have been used. This makes it difficult to address the
question of reddening for all fields and clusters in a consistent way.
We later consider the sensitivity of our conclusions to the adopted
reddening. Initially we consider available determinations in
the literature.

We have searched the literature for independent determinations of the
reddening. For two of the clusters and the SGR-I field observations in
the two HST U-passbands exist, which could in principle allow
derivation of the extinction directly. Unfortunately the clusters were
observed in F336W, which has a large red leak. This means that the
extinction has a non-linear dependence on the reddening which makes it
difficult to deredden the stars in the colour-colour diagram.

We use the extinction for a K5 spectrum as given in Table 12 of
Holtzman et al.\,(1995b) to calculate the reddening vector.  When
$E(B-V)$ is not directly available, we use the extinction law given in
Table 2 of Cardelli et al.\,(1989) to derive $E(B-V)$ (adopting
$R_V=3.1$). The extinctions and reddenings from the literature are
collected in Table \ref{reddenings}.

\begin{table}
\caption[]{Reddenings from the literature and extinctions, derived 
from Holtzman et al. (1995b)}
\begin{tabular}{llllllllllllllllll}
\hline\noalign{\smallskip}
Field & $E(B-V)$ & $A_{\rm F555W}$ & $E(V_{\rm F555W}-I_{\rm F814W})$\\
\noalign{\smallskip}
\hline\noalign{\smallskip}
SGR-I & 0.58 & 1.77 & 0.70 \\
BW    & 0.49 & 1.49 & 0.59 \\
NGC5927 &  0.46 & 1.41 & 0.56 \\
NGC6528 & 0.6 & 1.83 & 0.73 \\
NGC6553 & 1.0/0.8 & 3.03/2.45 & 1.20/0.50\\
\noalign{\smallskip}
\hline
\end{tabular}
\label{reddenings}
\end{table}

\subsection{Comparison with previous studies}

The same observations as we use for NGC6553 and NGC6528 have previously
been reported in Ortolani et al. (1995) and those for NGC5927 in
Fullton et al. (1996).  A comparison between our colour-magnitude
diagrams and theirs shows excellent agreement where even individual
stars may be identified. The comparison of both the general structure
of the colour-magnitude diagrams as well as of magnitudes of
individual stars give us confidence in our photometry for the other
fields, eg. MW, SGR-I.  Ortolani et al. (1995) did not
publish the colour-magnitude diagram for NGC6528 but the ridge-line.

\subsection{The Bulge fields}

\subsubsection{SGR-I}
\label{sgri_text}

SGR-I is a well known low-extinction area. Glass et al.(1995) adopt
$A_V=1.87$. They ascribe fairly large and uncertain errors to this
value. The cluster NGC6522 appears to have similar extinction to
SGR-I. Walker \& Mack (1986) found $A_V=1.78\pm0.10$. Using $R_V=3.1$
this corresponds to $E(B-V)=0.57$. Frogel et al. (1990) also derive
0.57 for an A0 star. Terndrup et al.(1990), from M giants, derive a
$E(B-V)=0.54$. The estimate is, essentially, based on the similarities
between NGC6522 and the SGR-I field. No HST observations in the
relevant passbands are available for NGC6522. We adopt $E(B-V)=0.58$.

\subsubsection{Baade's window at R.A.=18 03 11.7 $\delta$=--29 51 40.7}
\label{bw_text}

This field is situated in the low extinction area called Baade's
Window.  Stanek (1996) produced a detailed extinction map of this area
from OGLE data. The uncertainty, which is systematic, in this
determination mainly arises from the zero point. This grid has
30arcsec spacing, and we are cautioned that there may exist further
structure in the reddening below this spatial resolution. For the
position of the PC1 his map gives $E(V-I)=0.566$ and $A_V
=1.452$. Using Cardelli et al.\,(1989) this corresponds to
$E(B-V)=0.49$. Using Holtzman et al. (1995b) this translates to
$E(V_{\rm F555W}-I_{\rm F814W}) = 0.59$.

\subsubsection{Fields at (l,b)=(3\fdg6,-7\fdg0) and (2\fdg9,-7\fdg95)}

These fields are part of two parallel programs and we do not have
further information from observations of the extinction. Figure
\ref{cmd_606} however suggests that they suffer from similar
extinctions. Since $V_{\rm 606}$ is roughly 0.6 magnitudes brighter
than $V_{\rm 555}$ the turn-offs for these two fields appear very
similar in colour to those of SGR-I and Baade's window.  We primarily
use these observations for number counts in $I_{\rm 814}$.

\subsection{The globular clusters}

%%%%%%%%%%%%%%%%%%%%%%%%%%%%%%%%%%%
%%%%%%%%%%%%%% tables %%%%%%%%%%%%

\begin{table*}
\caption[]{Data for NGC5927 compiled from the literature.}
\begin{tabular}{llllllllllllllllll}
\hline\noalign{\smallskip}
  & Value & Ref. & Comment\\\noalign{\smallskip}
\hline\noalign{\smallskip}
Core radius & $0\farcm 42$& Harris (1996) &\\
$\Delta(m-M)$ & 16.10 & Zinn (1980) & 16.6 kpc\\
         & 14.52 & Djorgovski (1993)  & 8.0 kpc\\
 {[Fe/H]}&  $-0.16 $ & Zinn (1980) \\ 
	 & $-0.30$ &Zinn \& West (1984) \\
	 & $-0.32/-0.64$ & Rutledge et al. (1997) & Zinn \& West (1984) scale/Caretta\& Gratton (1997) scale\\
{[Fe/H]}$_{\rm 47 Tuc}$& $+0.57$ &  Cohen (1983) &  relative to 47 Tuc at --0.7 dex\\
Age & $10.9\pm2.2$ Gyr & Fullton et al. (1996) \\
	& $15$ Gyr & Samus et al. (1996) \\
E(B-V) & 0.48 & Zinn (1980) \\
       & 0.46 & Peterson (1993) \\ 
       & $(0.44) - 0.46$ & Sarajedini \& Norris (1994) & For several solutions for \\
       &                 &                             & reddening and metallicity\\
\noalign{\smallskip}
\hline
\end{tabular}
\label{ngc5927_tab}
\end{table*}

\begin{table*}
\caption[]{Data for NGC6528 compiled from the literature.}
\begin{tabular}{llllllllllllllllll}
\hline\noalign{\smallskip}
  & Value & Ref. & Comment\\\noalign{\smallskip}
\hline\noalign{\smallskip}
Core radius & $0\farcm 09$& Harris (1996) \\
 Distance & 7.5 kpc  & Ortolani et al.\,(1992) \\
$\Delta(m-M)$ & 16.4 & Zinn (1980) & 19.1 kpc\\
 {[Fe/H]}     & $+0.01$ & Zinn (1980)\\
              & $+0.29$ & Bica \& Patoriza (1983)\\ 
	      & $+0.12$ & Zinn \& West (1984) \\
	      & $-0.23$ & Armandroff \& Zinn (1988) & Integrated spectra IR Ca {\sc ii}\\
	      & high, sim to NGC6553& Ortolani et al.\,(1992)\\
              & $-0.23$ &Origlia et al.\,(1997) & IR abs. at 1.6 $\mu$m \\
 {[M/H]} & $+0.1/-0.4$ & Richtler et al.\,(1998)&Trippico isochrone/Bertelli isochrone\\
   $Z$ & $Z_{\odot}$ & Bruzual et al.\,(1997) & \\
 Age & 14 Gyr & Ortolani et al.\,(1992) & metallicity comparable to solar\\
	& $12\pm2$ Gyr & Bruzual et al.\,(1997) & \\
 E(B-V) & 0.55 & Ortolani et al.\,(1992) & NGC6553 as reference and $\Delta(m-M)_V= 14.39$\\
	& 0.62 & Bruzual et al.\,(1997) & \\
	& 0.56 & Zinn (1980) \\
  E(V-I)& 0.8/0.6 & Richtler et al.\,(1998)&Trippico isochrone/Bertelli isochrone\\
\noalign{\smallskip}
\hline\noalign{\smallskip}
\end{tabular}
\label{ngc6528_tab}
\end{table*}

%%%%%%%%%%%%%%%%%%%%%%%%%%%%%%%%%%%%%%%%%%%%
%%%%%%%%%%%%%%%%%%%%%%%%%%%%%%%%%%%%%%%%%%%%

\begin{figure}
\resizebox{\hsize}{!}{\includegraphics[angle=-90]{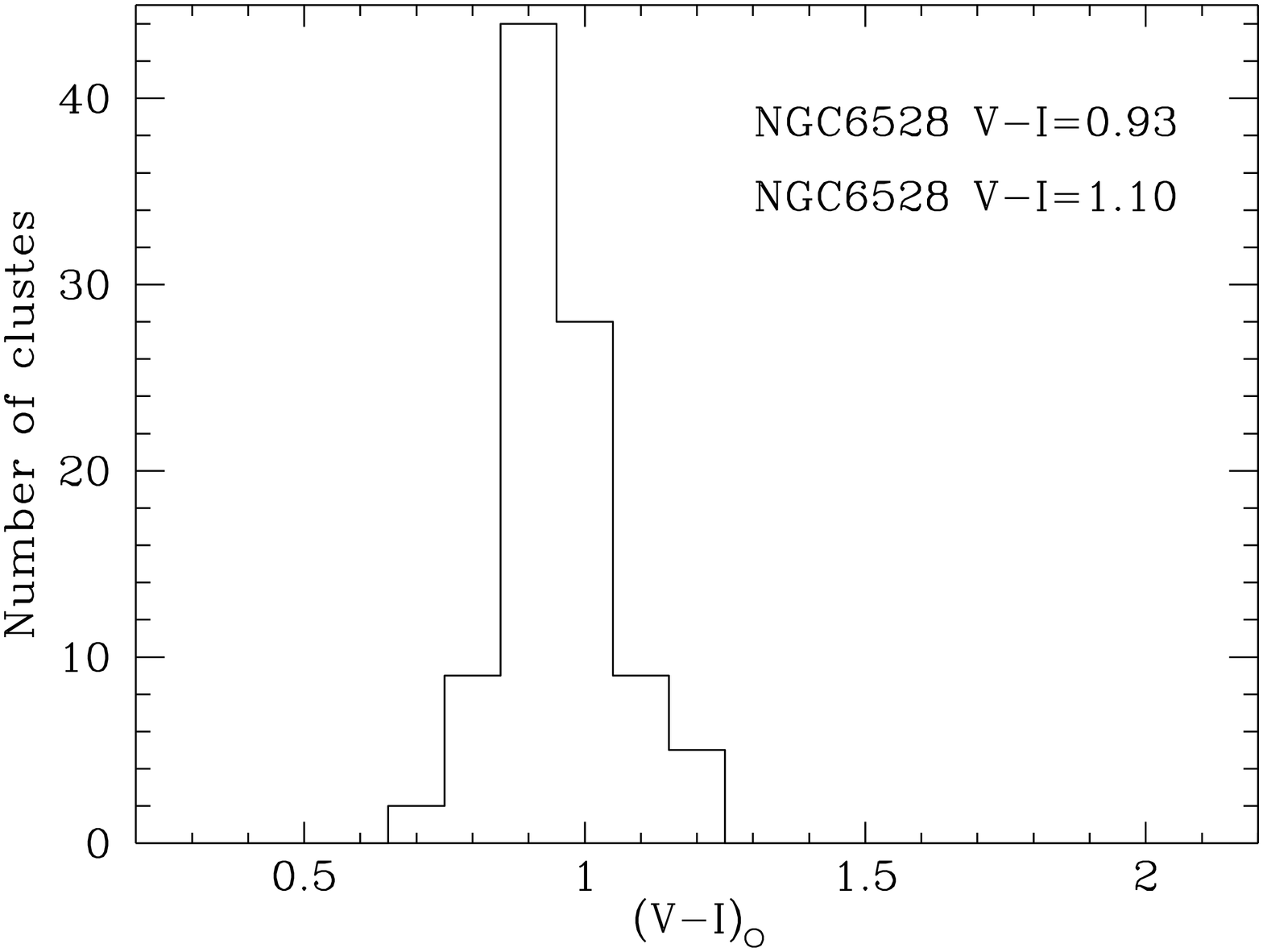}}
\resizebox{\hsize}{!}{\includegraphics[angle=-90]{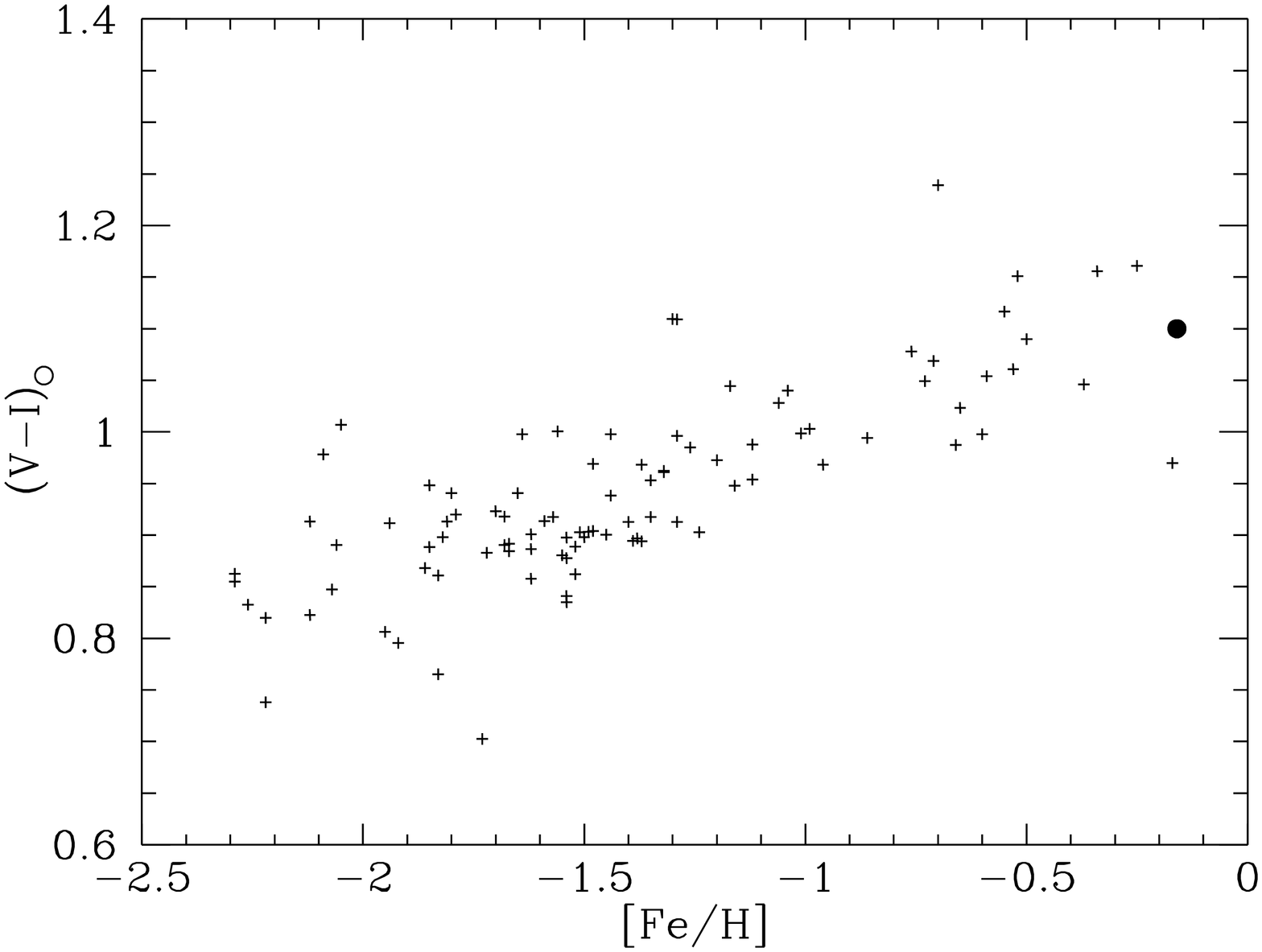}}
\caption[]{Histogram showing the distribution of integrated $(V-I)_o$
for all the clusters in Harris (1996) catalogue for which photometry
in all passbands are available. We have used the reddening quoted in
Harris to calculate the intrinsic colours. The catalogue values of
$V-I$ for NGC6528 and NGC6553 are given. Harris give $E(B-V)_{\rm
6528}=0.62$ and $E(B-V)_{\rm 6553}=0.84$. Note that if we used the
reddenings quoted in this paper the clusters would in fact become
intrinsically even less red.  The second panel shows the V-I colours as a
function of the [Fe/H] values quoted in Harris. The $\bullet$ in the
lower panel indicates the position of NGC6553 with the new iron
abundance from Cohen et al. (1999).}
\label{harris_histogram}
\end{figure}

\begin{figure}
\resizebox{\hsize}{!}{\includegraphics{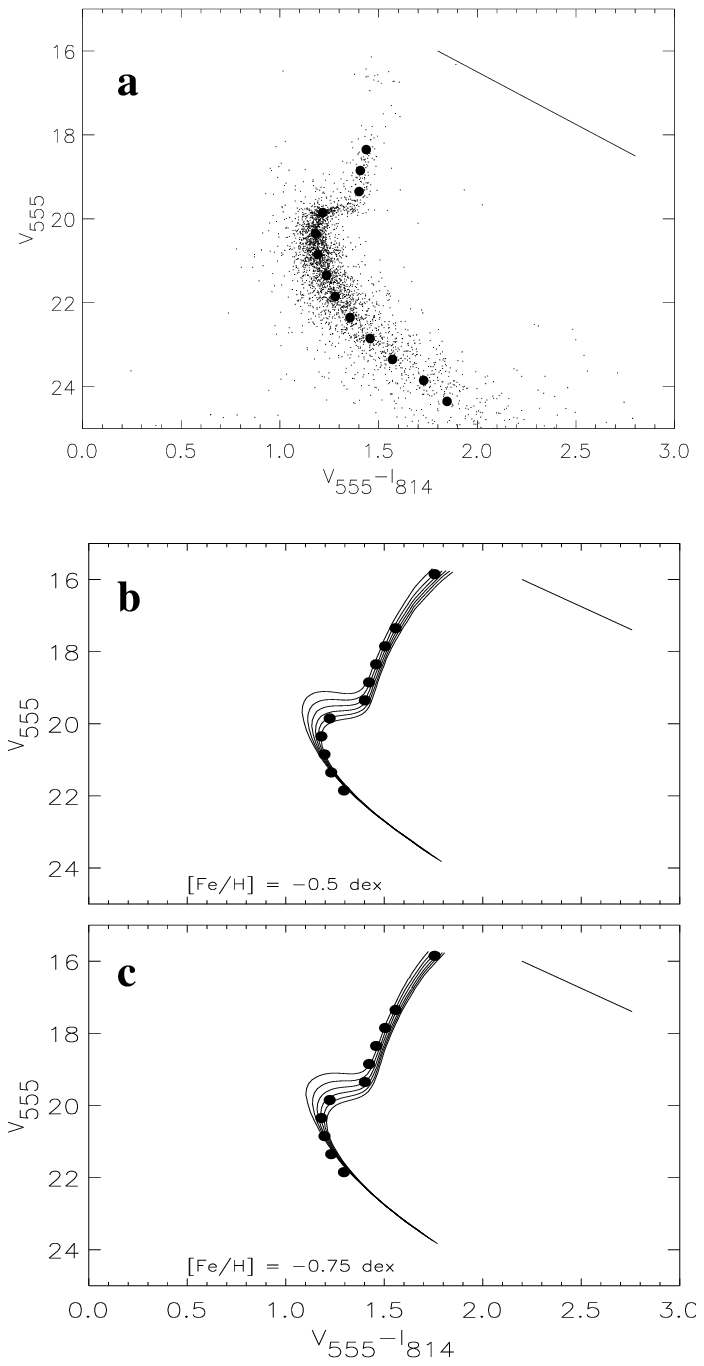}}
\caption[]{Colour-magnitude diagrams for NGC5927. In panel a
the long exposure. The ridge line used in Sect. \ref{metgrad.sec} 
is over-plotted
as heavy dots. In panel b and c are  the ridge line shown for the
 short exposure with
isochrones superimposed. Isochrones are for 8, 10, 12, 14, 16,
18 Gyr, Bertelli et al. (1994) and Guy Worthey (private
communication).  The isochrones have been moved to a distance of 8 kpc
and an extinction of $E(B-V)=0.46$.  Reddening vectors are plotted in
upper right hand corner. Metallicities as indicated}
\label{fit_ngc5927}
\end{figure}

%%%%%%%%%%%%%%%%%%%%%%%%%%%%%%%%%%%%%%%%%
%%%%%%%%%%%%%%%%%%%%%%%%%%%%%%%%%%%%%%%%%

Two of the clusters, NGC6528 and NGC6553, are situated between us and
the Galactic Bulge. The third, NGC5927, is also in the disk but well
away from the Bulge line of sight, hence we will only have
contamination from disk stars in this field.  All three are thought to
be among the more metal-rich globular clusters in the Galaxy (e.g.
Rutledge et al. 1997). The cluster kinematics have been interpreted as
evidence that the clusters are members of
what is called the bulge population of globular clusters (see Minniti
1996 and references therein). We noted above that such assignments do
not allow any deduction on the relative properties (ages, etc) of
field and cluster stars, given the lack of understanding of the
formation of either population.

We have searched the literature for data on the clusters. Several
studies are available, both in the visual and the IR as well as
spectroscopic studies of individual stars and of the integrated light
from the cluster, and are summarized in Tables \ref{ngc5927_tab},
\ref{ngc6528_tab} and \ref{ngc6553_tab}.   References
reporting a result are given in preference to later reviews.
It is important for our
study to understand the metallicities of the clusters and the fields,
at least on a relative scale. Large metallicity differences would
influence the conclusions from the number counts in
Sect. \ref{sect_age_bulge}.

We note, using the catalogue data by Harris (1996), that the integrated
colours for NGC6528 and NGC6553 show them to be ``normal''
clusters. This is illustrated in Fig.\ref{harris_histogram}, where it
is seen that extreme high metal abundances are unlikely, given their
integrated colours.

\subsubsection{NGC5927}
%%%%%%%%%%%%%%%%%%%%%%%  Figurer %%%%%%%%%%%%%%%%%%%%%%

This cluster is situated in the disk well away from the Bulge, and is
one of the most metal-rich disk globular clusters known. 
Fullton et al.\,(1996) present colour-magnitude diagrams from the same
HST/WFPC2 observations as we use. They derive a cluster age of
$10.9\pm2.2$ Gyr, using ${\rm [Fe/H]} = -0.24 \pm 0.06$ dex. This makes the
cluster 3-5 Gyr younger than many other disk clusters. The lower
metallicity found from Ca{\sc ii} infrared triplet lines by Rutledge
et al. (1997) would imply a higher age. Samus et al. (1996) presented
the first deep ground-based  $BVI$ photometry for the cluster from
which they derived an age of 15 Gyr assuming ${\rm [Fe/H]} =-0.49$ dex.

In Fig. \ref{fit_ngc5927} we show the ridge-line of NGC5927 with
theoretical isochrones for 8-18 Gyr over-plotted. The isochrones have
been moved to a distance of 8 kpc and an extinction of $E(B-V)=0.46$
(see Table \ref{ngc5927_tab}).  From this we infer that $[{\rm
Fe/H}]=-0.64$ dex, as derived by Rutledge et al.\,(1997), is a good
fit to the data.  Also that a very young age is not likely, but rather
$12-14$ Gyr.

\subsubsection{NGC6528} 
\label{sec_ngc6528}

NGC6528 is situated in the line of sight towards Baade's Window. It
has been the subject of two separate studies (Ortolani et al.
1992 and Richtler et al. 1998), as well as figuring in a number of
studies of other metal-rich globular clusters (Cohen \& Sleeper 1995,
Ortolani et al. 1995, Kuchinski et al.1995, Bruzual et al.1997).  To
our knowledge no spectroscopic abundances have been reported.

From the similarities in field and cluster colour-magnitude diagrams,
Ortolani et al.\,(1992) concluded that not only is NGC6528 projected
onto the Galactic Bulge but the stellar population is
indistinguishable from that of the bulge population.  They noted a
strongly curved red giant branch, interpreted as evidence for high
metallicity, and a tilted horizontal branch. Richtler et al. (1998)
could not identify the cluster red giant branch directly in their
colour-magnitude diagram due to the combination of the Bulge population
in the observed area and cluster AGB stars. 
The initial tilt of the horizontal branch was
significantly reduced by carefully selecting bulge stars and
discarding the field stars. By investigating the member-ship of the
very red giants, $V-I > 3.5$, they conclude that the cluster is indeed
situated in front of the general bulge field population and not embedded in
it.

Our data do not reach bright enough magnitudes to study the horizontal
branch in detail, however, from number counts in
Sect. \ref{sect_age_bulge} we conclude that our colour-magnitude
diagram is significantly contaminated by Bulge stars. Thus, while
NGC6528 is possibly not a clean fiducial cluster, it is ideal for our
present age-test experiment (see also Sect. \ref{sec_disc}).

\subsubsection{NGC6553}

NGC6553 is perhaps the best studied cluster of the three.  Ortolani et
al.\,(1990) published the first detailed study of this cluster, which
is situated at a projected distance of $\sim 6\degr$ from the galactic
centre and roughly at a distance of 5 kpc from us. Being some 10-20
scale lengths from the Galactic centre, this makes it a
marginal `bulge' cluster. However it is often included in such studies
(eg Barbuy et al. 1998).  Here we will use its colour-magnitude
diagram to constrain the number of young stars in the Galactic Bulge,
Sect. \ref{sect_age_bulge}.  However, the colour-magnitude itself
merits a further discussion here.  The work of Ortolani et al has been
followed up in the infra red by Davidge \& Simons (1994), and by
Guarnieri et al.\,(1997) who present a study based on both
optical and infrared observations. The aim with this study is to
provide a template for infra red studies of the galactic bulge
population. 

\subsubsection{Differential reddening towards NGC6553?}
\label{sec_diff_red}

%%%%%%%%%%%%%%%%%%%%%%%%%%%%%%%%%

\begin{figure*}
\resizebox{12cm}{!}{\includegraphics{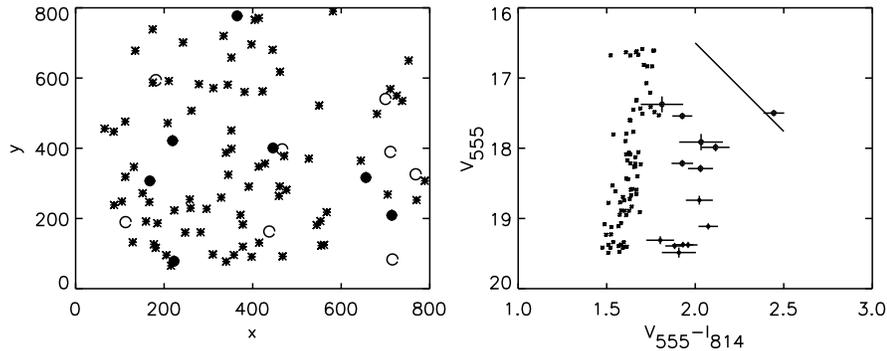}}
\hfill
\parbox[b]{55mm}{
\caption[]{NGC6553. a. The position on the PC1 chip for the stars on
the two giant branches. b.  Colour-magnitude diagram for the same
stars. The direction of the reddening is indicated by a solid
line. Open circles denote stars with $V_{\rm 555} < 18.5$ and filled
circles stars with $V_{\rm 555} > 18.5$. Error bars, as given by {\sc
allstar}, are over plotted for these stars.}
\label{map_sel_ext}}
\end{figure*}	

Our colour-magnitude diagram, in Fig. \ref{cmd_shallow}, shows some
intriguing and previously not discussed patterns.  At $V \sim 21.5 $ an
apparent second turn-off is seen and more or less parallel to the red of the
red giant branch is a second sequence of stars.  This sequence is
actually visible also in the colour-magnitude diagram in Ortolani et
al.\,(1995), however the apparent turn-off is not. Additionally, this
cluster has a well-known `tilted' horizontal branch, with the HB slope
lying close to that of the reddening vector, and has a broad
main-sequence turn-off.

Can the ``extra'' giant branch, and the various other anomalies be due to
differential reddening of cluster stars? Ortolani et al. (1995) found the
reddening to be variable over the PC1, $\sim 0.2$ magnitudes across
the field.   The simplest possible test is to see if the CMD anomalies
are restricted to one patch of the field that has higher
reddening. 
In Fig. \ref{map_sel_ext} we show the positions of the stars in the
``extra'' giant branch. It is clear that the stars are evenly
distributed over the image and that their positions in the
colour-magnitude diagram cannot therefore be due to a simple selective
extinction effect. Plausible astrophysical explanations all have difficulties:
the obvious explanation, that this is the background
Galactic bulge, with some additional reddening, is very difficult to
make consistent with the data.  An alternative, though also
speculative, explanation is that the second giant branch is part of
the Sgr dSph galaxy (Ibata et al. 1995). The absence of any such
features in Fig.\ref{cmd_606} for fields closer to the centre of Sgr
requires a very non-uniform surface density in the dwarf
galaxy. However, even with this explanation, further reddening beyond
NGC6553 is required to move it to such red colours. 

An explanation of the colour-magnitude data for NGC6553 remains
difficult. Reddening which is both patchy and mixed through the
cluster remains feasible. A detailed HST (WFPC plus NICMOS) study is
required, and is underway. It will be reported elsewhere  (Beaulieu et
al. 1999, in prep.). In the interim, some reserve in deductions from
this cluster is advised.

\subsubsection{Spectroscopic abundances}

In addition to the metallicities based on photometry and spectroscopy
of the Ca {\sc ii} infra red triplet, summarized in Table
\ref{ngc6553_tab}, two studies have obtained high-resolution abundance
data for a handfull of stars in the cluster.  Barbuy et al. (1997)
completed the first detailed abundance study and derive Mg, Ti, Si, Ca
and Eu abundances as well as Fe abundances for 3 very cool
stars. Preliminary results are ${\rm [Mg/Fe]} \simeq +0.15$, ${\rm
[Ti/Fe]} \simeq +0.3$, ${\rm [Si/Fe]} \simeq +0.6$, ${\rm [Ca/Fe]}
\simeq 0.0$ and ${\rm [Eu/Fe]} \simeq +0.3$. The abundance ratios for
Mg, Ti, Ca and Eu are similar for those found in metal-rich dwarf
stars in the solar neighbourhood (Feltzing \& Gustafsson 1998 and
Feltzing 1999) while the cluster appears overabundant in Si.  This
could point to a rapid star formation history, but any interpretation
is contradicted by the low Ca abundance, which is not consistent with
the other alpha-elements.

Cohen et al. (1999) find that five horizontal branch stars have a mean
[Fe/H] of $-0.16$ dex, which is comparable to the mean abundance in the
Galactic Bulge as found by McWilliam and Rich (1994). The horizontal
branch stars are preferable to use because their spectra are less
crowded and  abundances are therefore more readily extractable then in
the very crowded spectra used by Barbuy et al. We will therefore adopt
this new high [Fe/H]. This is also consistent with the early findings
by Cohen (1983) that an underestimate in $E(B-V)$ of as little as
$0.^m05$ corresponds to an underestimate in [M/H] of 0.2 dex.  

\begin{table*}
\caption[]{Data for NGC6553 from the literature.}
\begin{tabular}{llllllllllllllllll}
\hline\noalign{\smallskip}
  & Value & Ref. & Comment\\
\noalign{\smallskip}
\hline\noalign{\smallskip}
Core radius & $0\farcm55$& Harris (1996) & \\
 Distance & 4.9 kpc  & Ortolani et al.\,(1990)\\
$\Delta(m-M)$ & 16.4 & Zinn (1980) & 19.1 kpc\\
              & 13.6 $\pm 0.25$ & Guarnieri et al.\,(1992) & 5.25 kpc\\
 {[Fe/H]}   & $+0.26$ &  Zinn (1980) \\
        & $+0.47$ & Bica \& Pastoriza (1983)& For a discussion of probable error sources \\
        & & & see Barbuy et al.\,(1992) (CNO excess) \\
        & $-0.7$ & Pilachowski (1984) & Spectroscopy of 1 star\\
        & $-0.41$ & Webbink (1985) \\
	& $-0.2^{+0.2}_{-0.4}$& Barbuy et al.\,(1992) & Spectroscopy of star III-17\\
	& $\geq -0.4$  & Davidge \& Simons (1994) & IR photometry\\
	& $-0.29$ & Zinn \& West (1994) \\
        & $-0.55$ & Barbuy et al.\,(1997), Barbuy et al.\, (1999) & Spectroscopy of 3 giant stars \\
        & $-0.33$ & Origlia et al.\,(1997) & IR abs. at 1.6 $\mu$m \\
        & $-0.60$ & Rutledge et al.\,(1997) & IR Ca {\sc ii} triplet \\
        & $-0.16$ & Cohen et al. (1999) & from 5 horizontal branch stars \\
{[Fe/H]}$_{\rm 47 Tuc}$& $+0.24/+0.37$ & Cohen (1983) & Spectroscopy 5 stars/2 most metal-rich stars, \\
        &               &       &  relative to 47 Tuc at --0.7 dex\\
	& $+0.10$ & Cohen \& Sleeper (1995) & relative to 47 Tuc at --0.71 dex \\
$Z$     & $Z_{\odot}$ & Bruzual et al.\,(1997) & \\
 Age    &  & Ortolani et al.\,(1990)& between 47Tuc and Pal~12\\
	& $12\pm2$ Gyr & Bruzual et al.\,(1997) & \\
 E(B-V) & 0.78 & Zinn (1980) \\  
        & $\leq 1.0$ & Ortolani et al.\,(1990)\\
	& 0.7 & Guarnieri et al.\,(1997)\\
\noalign{\smallskip}
\hline
\end{tabular}
\label{ngc6553_tab}
\end{table*}
%
%________________________________________________________________
\section{The age of the Galactic Bulge}
\label{sect_age_bulge}

\begin{table*}
\caption[]{Number of stars in ``boxes''. The first column identifies
the fields and clusters, the second the designation, i.e. MS (main sequence), 
TO (turn off), and Disk, for the box, the third gives the range in $V_{\rm
555}$ and the fourth gives the range in $V_{\rm 555}-I_{\rm 814}$ used for 
defining the box, then follows the number of stars counted in each box. 
The last three columns  give the relative numbers in the different boxes.}
\begin{tabular}{lrrrrrrrrrrrrrrrrrrrrrr}
\hline\noalign{\smallskip}
Field & &V & V-I & \#  & Disk/MS & Disk/TO & TO/MS \\
\noalign{\smallskip}
\hline\noalign{\smallskip}
SGR-I   &Disk & 18-20       & 0-1.6  & 54    &0.13      &0.20 &\\
(d=3deg)  &TO   & 20-21       & 0-1.6  & 268   &          & &0.07\\
        &MS   & 21-22       & all    & 404   &          & &\\
BW      &Disk & 18-19.7     & 0-1.35 & 34    &0.12      &0.18 &\\
(d=4deg)   &TO   & 19.7-20.7   & 0-1.4  & 188   &          & &0.07\\
1000s        &MS   & 20.7-21.7   & all    & 276   &          & &\\
BW      &Disk & 16(18)-19.7 & 0-1.35 & 47(39)&0.15(0.13)&0.24(0.20) &\\
40s     &TO   & 19.7-20.7   & 0-1.4  & 193   &          & &0.06\\
        &MS   & 20.7-21.7   & all    & 307   &          & &\\
NGC6528 &Disk & 18-20       & 0-1.5  & 50    &0.11      &0.18 &\\
(d=5deg)        &TO   & 20-21       & 0-1.5  & 278   &          & &0.06\\
        &MS   & 21-22       & all    & 473   &          & &\\
NGC6553 &Disk &   -19.5     & 0-1.44 & 26(31)&0.04(0.05)&0.07(0.08) &\\
(d=6deg)        &TO   & 19.5-20.5   & 0-1.44 & 397   &          & &0.06\\
        &MS   & 20.5-21.5   & all    & 667   &          & &\\
NGC5927 &Disk & 18-19.6     &  0-1.3 & 52    &0.06      &0.06 &\\
(d=35deg)    &TO   & 19.6-20.6   & 0-1.3  & 809   &          & &0.09\\
        &MS   & 20.6-21.6   & all    & 889   &          & &\\
\noalign{\smallskip}
\hline
\end{tabular}
\label{number_counts}
\end{table*}

\begin{figure}
\centerline{\resizebox{7cm}{!}{\includegraphics{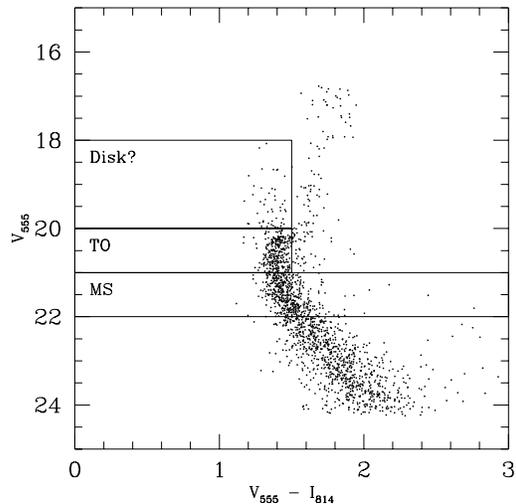}}}
\caption[]{Definition of windows for NGC6528, Table \ref{number_counts}.}
\label{windows.fig}
\end{figure}

\begin{figure}
\centerline{\resizebox{7cm}{!}{\includegraphics{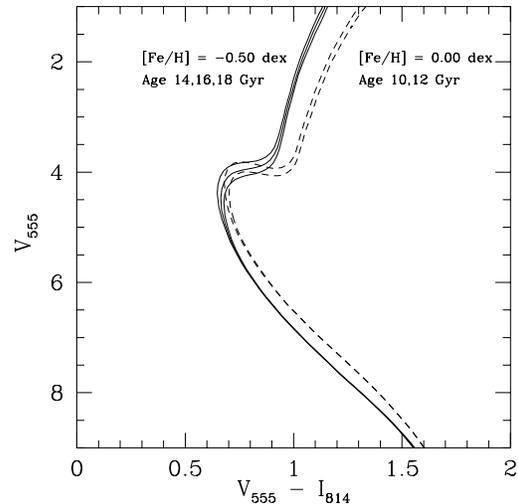}}}
\caption[]{Illustration of the age-metallicity degeneracy around
the turn-off. Isochrones are from Bertelli et al. (1994), in the HST
passbands courtesy of G. Worthey. }
\label{comp_iso}
\end{figure}

In our colour-magnitude diagrams in Fig \ref{cmd_shallow} and
\ref{cmd_deep} there appear to be many stars in the ``young stars''
and ``blue straggler'' region, brighter and bluer than the main
main-sequence turn-off. Because the turn-off region is sensitive to
age (Fig. \ref{comp_iso}) is this evidence for a young stellar
population?  The turn-off region in these diagrams is also sensitive
to foreground contamination, and to bulge blue stragglers.  
Rather than simply assuming the nature
of the stars around the turn-off in the colour-magnitude diagrams of
Bulge fields, foreground or a substantial young Bulge population, we
test the possibility that they are foreground disk contamination. This
is done by quantifying their spatial distribution, since foreground
disk stars will be distributed on the sky differently than are bulge
stars, of whatever age. The key to this experiment is our
consideration of fields in a variety of directions, and in particular
use of the cluster NGC5927, which is at longitude $34 \degr$, 
far from the bulge.

We defined three ``windows'' in our colour-magnitude diagrams in which
we performed number counts. The location of the windows was decided
upon by inspecting the colour-magnitude diagrams of SGR-I and
NGC6528. We defined three windows; one corresponding to stars above
the turn-off (young bulge and/or foreground disk and/or blue
stragglers),  one at the turnoff
of an old population and one on the bulge main sequence, as
illustrated in Fig. \ref{windows.fig}. The colour and magnitudes
limits were then adjusted for each field and cluster to take distance
and reddening differences into account, following Table
\ref{number_counts}.

We first compare the counts for the two clusters, NGC6553 and NGC5927,
one near the bulge, one far out in the disk. For these two clusters,
The relative number of stars that are either young or are foreground
disk stars and stars that are thought to belong to the Bulge, called
``Disk/MS'' in the table, is constant between the fields, while the
absolute number is roughly constant.  This is direct evidence that the
counts in the disk box are dominated by true disk stars and not true
bulge stars. 
Further direct evidence that the ``disk'' stars are disk, is
the near constancy of their surface density in the inner bulge fields.

In this part of the diagram the star numbers do not
change with $l,b$ as they would if we were seeing a young Bulge
population. The COBE Bulge has a scale length of $\sim 1^o$ (Binney,
et al. 1997) and hence the number counts would change
significantly over the area studied.

The ratio of disk to main sequence and turn-off stars for NGC6528 is
different to that for the other two clusters, suggesting that for this
cluster the colour-magnitude diagram is significantly affected by
field contamination, perhaps explaining some of the apparent anomalies
noted in earlier analyses of these HST images (Ortolani et al. 1995).

%%%%%%%%%%%%%% Figures %%%%%%%%%%%%%%%%%%%%%

\begin{figure}
\resizebox{\hsize}{!}{\includegraphics{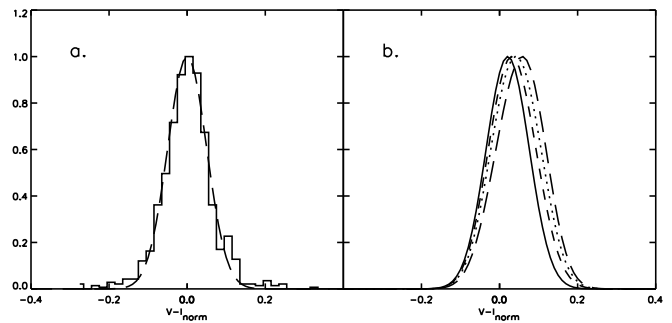}}
\caption[]{In panel a. we show the colour distribution for stars in
NGC5927 with $20.2 \leq V_{\rm 555} \leq 21.2$. The histogram has been
moved so that its mean colour ($V-I_{\rm norm}$) is zero. To this
histogram a Gaussian is fitted, shown by the dashed line. This Gaussian
defines  the colour distribution in a single stellar population,
convolved with our photometric errors.  A colour
difference of 0.07 in magnitude corresponds to a change in [Fe/H] of
0.3 dex (at -0.3 dex and 0.0 dex).  Two Gaussians separated in colour
by the equivalent of a 0.3 dex metallicity difference
are added to make up  a reasonable representation of the colour
distribution in Baade's window (see text). This is plotted in b. To
the Gaussian representing Baade's window a third population at +0.3
dex is added in different proportions, to illustrate the effect on the
observed colour distribution of a very metal-rich population. The
three curves are:
short dashed line 50\% stars added, dotted line 100\% stars added and long
dashed line 200\%.}
\label{fe_vs_col}
\end{figure}

\begin{figure*}
\resizebox{12cm}{!}{\includegraphics{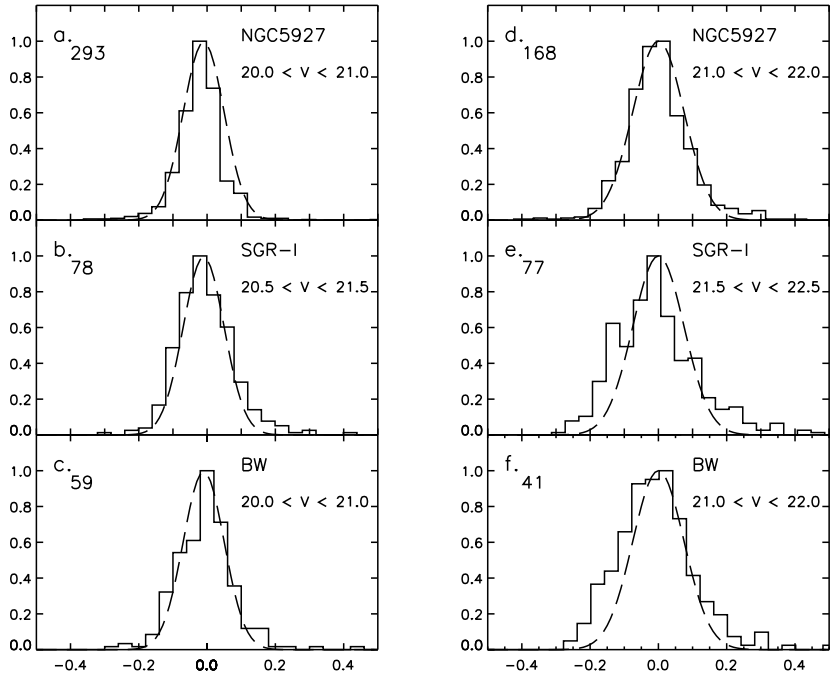}}
\hfill
\parbox[b]{55mm}{
\caption[]{Normalized histogram of the colour distribution of 
stars in magnitude slices for NGC5927, SGR-I
and the deep exposure of Baade's window. The slices are made in
V$_{\rm 555}$, with magnitude ranges appropriate for reddening and
distance in each line of sight. The magnitude ranges are:
panels a and c are for 20-21, b for 20.5-21.5,
d and f for 21-22, and e for 21.5 -22.5. }
\label{slices_deep}}
\end{figure*}

%%%%%%%%%%%%%%%%%%%%%%%%%%%%%%%%%

\begin{table}
\caption[]{Number counts in $I_{\rm 814}$ for the four fields. Their
relative numbers are compared with simple predictions using the E2 
model of the Galactic Bulge density distribution from Dwek et al. (1995).}
\begin{tabular}{lrrrrrrrrrrrrrrrrrrrrrr}
\hline\noalign{\smallskip}
Field & $\Delta I_{\rm 814}$ & \#  & rel. SGR-I & Model \\
\noalign{\smallskip}
\hline\noalign{\smallskip}
SGR-I & 20.0-21.0 & 536 & -- & --\\
BW & 19.5-20.5 & 372& 0.69 & 0.62\\
   & 20.0-21.0 & 328 & 0.61 \\
u811 & 20.0-21.0 & 52 & 0.10 & 0.11\\
Deep & 20.0-21.0 & 44 & 0.08 & 0.09 \\
\noalign{\smallskip}
\hline
\end{tabular}
\label{number_counts_sim}
\end{table}

The spatial distribution of the number counts in $I_{\rm 814}$ in the
four deep fields are consistent with results from simple simulations
using the E2 model  of the Galactic Bulge in Dwek et al.\,(1995)
(Table \ref{number_counts_sim}).  We present two counts for Baade's
window. This field has the lowest  extinction and we would expect the
counts between $19.5 \la I_{\rm 814} \la 20.5$ to be comparable with
the counts in SGR-I between 20.0 and 21.0. However, for comparison we
also give the counts in the same apparent magnitude  bin for Baade's
window. As  can be seen the difference is small, but consistent with
Baade's window having an extinction $\sim$ 0.3 magnitudes less than that
towards SGR-I.

We conclude that the apparent evidence for a significant young Bulge
population in our colour-magnitude diagrams is an artifact of disk
contamination. Correcting the observed colour-magnitude diagrams,
using the results for NGC5927, removes all the stars above and to the
blue of the turnoff. That is, there is no evidence in these HST/WFPC2
data for a significant age range in the bulge population. 

This tight limit, extending the previous results of Ortolani etal that
the bulk of the bulge is old, is somewhat surprising. What happens to
all the young stars forming in the inner disk?
We emphasize however, that available limits still allow an age range of
several Gyr, especially so, as discussed below, if there is an
age-metallicity relation in the bulge stars.

\subsection{Can we resolve the metallicity distribution function?}

We now consider if we are able to use the apparent width of the
main-sequences in the colour-magnitude diagrams to constrain the width
of the  stellar metallicity distribution function. Recall, from Fig. 
\ref{comp_iso} that a younger metal-richer population can hide in the 
turn-off region. It should, however, show up on the main sequence, if 
the statistics are good enough.

In Fig.\ref{fe_vs_col} we show a slice histogram for NGC5927 for all
stars between $V_{\rm 555}$ 20.2 and 21.2 and a Gaussian fitted to
it. We use this Gaussian as a representation of the apparent colour
distribution including measurement errors of a single stellar
population in our data.  A difference of 0.07 in colour corresponds to
a change in [Fe/H] of 0.3 dex.  Two Gaussian separated by 0.3 dex are
added to make up the  Gaussian drawn with a full line. This should be
a reasonable representation of the colour distribution in Baade's
window (see text). To the Gaussian representing Baade's window we add
a third population at +0.3 dex. As is clear from the figure, to be
able to resolve a metal-rich extra population,  that population has to
be large in numbers relative to the metal-poor one, and enough stars have
to be observed so that the histograms can be constructed with small
enough bins ($\sim$ 0.02). This is not the case here.

In Fig. \ref{slices_deep} 
the histograms have been moved so that the mean colour for each slice
is centred at 0.  We have fitted a Gaussian to the two histograms for
NGC5927. This Gaussian is reproduced in the panels showing the
histograms for SGR-I and Baade's window.  From this we conclude that
there is some evidence for a larger spread in  $V-I$ in the fields
than in the cluster. In particular, the slice at a fainter $V_{\rm
555}$  show a larger spread towards the blue, for the two field
populations than for the cluster, consistent with a broad metallicity
distribution function with a tail to metal-poor stars. The metal-rich
(red) sides of the histograms are however too poorly determined for
useful conclusions.

\section{Is there a detectable metallicity gradient?}
\label{met-grad}

In order to check a possible metallicity gradient, we compare the
colour-magnitude diagrams of the Baade's window and SGR-I field.  This
is a particularly robust method, since we are not sensitive to
incompleteness at the faint end of the ridge lines, the
colour-magnitude diagrams being comparably complete.

\subsection{The metallicity in Baade's window}
\label{metgrad.sec}

%%%%%%%%%%%%%%%%%%%% Figurer %%%%%%%%%%%%%%%%%%%%5

\begin{figure}
\resizebox{\hsize}{!}{\includegraphics{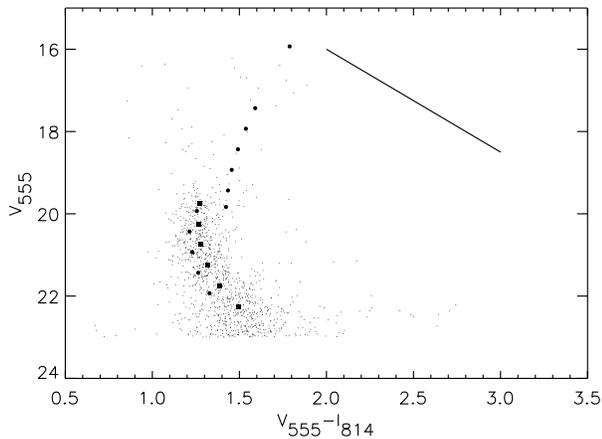}}
\caption[]{The CMD for Baade's Window is shown, together with its own
ridge line (fainter than $V_{\rm 555} \sim 20$, open squares). The
solid points ($\bullet$) are the ridge line constructed from the
colour-magnitude diagram of NGC5927, shifted along the reddening line
appropriately. }

\label{bw40_ngc5927}

\end{figure}

\begin{figure*}
\resizebox{12cm}{!}{\includegraphics{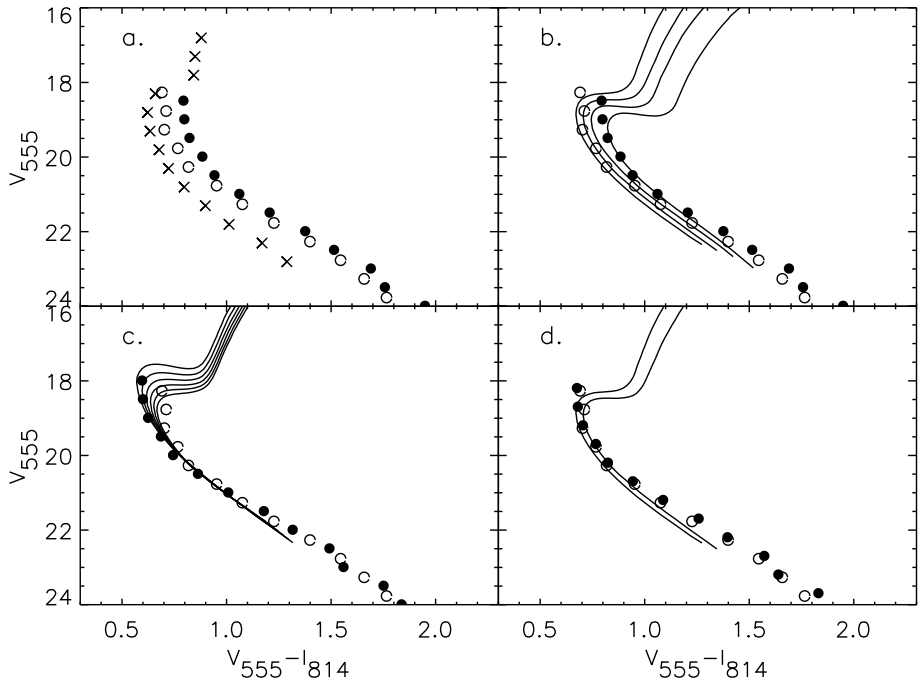}}
\hfill
\parbox[b]{55mm}{
\caption[]{Each of the 4 panels shows the comparison of the ridge
lines for SGR-I (filled circles) and Baade's window, (open
circles). Panel (a) presents the ridge lines as observed, corrected
according to the reddenings cited in the literature. This panel also
shows the ridge line for NGC5927, ($\times$). In (b) we also over plot
a set of 14 Gyr old isochrones for $[Fe/H] = -0.5, -0.25, 0.0, +0.25 $
dex, illustrating the apparent large abundance gradient implied by
these adopted reddenings. In panel (c) the ridge lines as observed are
moved to optimize agreement between the lower part of the two main
sequences. Isochrones for $[Fe/H] = -0.25 $ dex and 8, 10, 12, 14, 16,
and 18 Gyr (Bertelli et al. 1994 and Worthey private communication)
are over-plotted, illustrating the large age gradient implied by this
method.  In panel (d) the ridge lines are shifted to optimize an
overall fit of the morphology. 14 Gyr isochrones for [Fe/H]= -0.5 and
-0.25 dex are over-plotted to indicate the sensitivity to abundance
gradients in this case.}
\label{comp_ridges}}
\end{figure*}

%%%%%%%%%%%%%%%%%%%%%%%%%%%%%5
%%%%%%%%%%%%%%%%%%%%%%
Figure \ref{bw40_ngc5927} shows the colour-magnitude diagram from the
short exposure of Baade's window with its own ridge line, and also the
ridge line derived from the NGC5927 short exposure data. The ridge
line of the cluster has been moved along the reddening line to account
for the differences in reddening for the two fields. We observe that
the both the main sequence and the giant branch of the field
population(s) are redder than those of the cluster. From this we
conclude that the mean metallicity of Baade's window is $\sim0.3$ dex
higher than that of the cluster. This is in good agreement with
results from spectroscopic studies.

\subsection{The Bulge metallicity gradient}

The most robust way to compare colour-magnitude diagrams is to
construct a ridge line for each field.  The ridge lines can then be
moved to compensate for differential reddening effects, and compared
to appropriate isochrones. This is illustrated in
Fig. \ref{comp_ridges}. We see, in panels (a) and (b), that adopting
literature values for the reddening, and attributing all the resulting
difference between the ridge lines to metallicity, implies a
difference of 0.5 dex in metal abundance, which corresponds to 3.2
dex/kpc, assuming a distance of 8 kpc to the Galactic Bulge.

However, the relative reddening between the fields may well be in
error.  We thus test two further hypotheses. First, that there is no
significant metallicity gradient, but there is an age gradient. This
is shown in Fig.\ref {comp_ridges} panel (c), where one sees that an
age gradient of $\sim 6$ Gyr over 0.14 kpc is required, a result we
consider implausible.

We therefore try a further experiment, assuming the reddening is
unknown, and is a free parameter to be determined. In this case we can
derive a lower limit to any real age or abundance gradients, if we
conservatively assign as much as is possible of the differences
between the colour-magnitude diagrams to reddening, while fitting the
two ridge lines using their overall morphology. This is shown in
panel (d) of Fig. \ref{comp_ridges}.  No significant residual
systematic difference between SGR-I and Baade's window remains. 

Using the isochrones by Bertelli et al.  (1994) our best estimate of
any allowed difference is $\la 0.25$ dex. We conclude that there is no
strong evidence for an abundance gradient between Baade's window, at
projected distance from the centre of 550 pc, and SGR-I, at projected
distance 412 pc, assuming a distance of 8 kpc for the Bulge. The upper
limit on the amplitude of any abundance is $\la 0.2$ dex,
corresponding to $\la 1.3$ dex/kpc.  This value may be compared with
the recent suggested detection, by Frogel et al. (1999),
of a gradient of amplitude 0.5dex/kpc.

\section{Discussion}
\label{sec_disc}

Our conclusion that the Galactic Bulge is predominately old is in
conflict with those of Vallenari et al. (1996) and Holtzman et
al. (1993). Both these studies found evidence for a substantial (up to
30\%) young stellar population in Baade's window. Also Kiraga et
al. (1997), from OGLE data, concluded that the disk stellar population
in Baade's window is old, while the true Bulge stars have a bluer
turn-off implying a younger age than 47 Tuc. These results are based
on Baade's window only. The strength of our study is that it utilizes
five different line of sight. Our use of data for the cluster NGC5927
is of considerable value, since this cluster is situated in the disk
well away from the sight line towards the Galactic Bulge. Accordingly
the contamination present in this colour-magnitude diagram must arise
from foreground disk stars.  We show that the relative and absolute
number counts in this cluster show the same pattern as those in the
Bulge fields, showing that the stars previously identified with a
substantial (or exclusively) young stellar population in the Galactic
Bulge are in fact foreground disk stars.  Holtzman et al. (1993) note
that their interpretation of the data would likely change
substantially if for example the reddening estimates are in error.
Our results are mainly robust against such errors.  Our results are
also in accordance with ground-based studies, especially in the IR,
which have found no evidence for a substantial young stellar
population in the Galactic Bulge (e.g. Terndrup 1988, Tiede et
al. 1995, Frogel et al. 1990).

The first HST analysis of the stellar populations in Baade's Window
which deduced an old age for the bulk of the bulge stars is that of 
Ortolani et al. (1995). We use the same HST data, and some other
archive data, to provide a direct test of a critical assumption
underlying that study, and to extend the analysis to search for
gradients in metallicity and/or age. By further removing from the
analysis the assumption that the `bulge' globular clusters are a true
tracer of the age and metallicity of the field stars, we restrict the
analysis to the mean bulge star: the metal rich bulge stars, those
above about solar abundance, may be substantially younger than the
bulk of the bulge.

Ortolani et
al. (1995) observed two `bulge' globular clusters, NGC6553 and
NGC6528, with HST/WFPC2. By comparing ridge lines of these globular
clusters with that of the metal-rich globular cluster 47 Tuc they
concluded that the `bulge' clusters have ages comparable to the halo
globular clusters. They then compared the V-band luminosity function
of all stars observed in Baade's Window with the V-band luminosity
function observed for NGC6528, concluding that ``the main sequence
luminosity functions also coincide with extremely high precision in
the brighter, age-dependent part that is less affected by
incompleteness and field contamination (that is, $19.5 \la {\rm V} \la
20.5$)''. As is illustrated clearly in our colour-magnitude data for
NGC6528 this part of the apparent magnitude
range is indeed heavily contaminated by foreground disk stars, which
appear as if they are young bulge stars. Why did these stars not
vitiate the Ortolani etal analysis? The explanation is that analyses
based only on luminosity functions are valid if and only if the field
contamination is exclusively foreground disk, and that there is no
significant young bulge population. In that case, the cluster
luminosity function is in error in just the same way as is the field
data, and so the two errors compensate. In fact, the luminosity
function method can be justified only {\sl post hoc}, after an analysis
of the type reported here. We confirm that Ortolani et al. (1995) were correct
in that assumption.

The observational study of the inner bulge is further confused by the
distinct possibility that the very centrally-concentrated ``infra-red
bulge'' seen by IRAS and COBE is not related in a simple way, if at
all, to the larger ``optical bulge'' studied further from the centre
(Ibata \& Gilmore 1995a, 1995b; Wyse et al. 1997, Unavane
\& Gilmore 1998, Unavane et al. 1998). Recent studies of external
galaxies (Carollo 1999) show that central nuclei are common in bulges,
and also that bulges have a diversity of properties.
Optical studies in the Milky Way have been
restricted to Baade's window and beyond, $\geq 4$ COBE scale lengths
from the centre. Here we compare Baade's window with the low
extinction window called SGR-I, roughly at the limit of optical
observations, at $b=2\fdg6$, testing to see if the inner and outer
Galactic Bulge have the same stellar population(s).

We deduce an upper limit to a metallicity gradient of $\la 1.3$
dex/kpc.  Is such an amplitude surprising?  Let us consider two
distinct possibilities which show that such a result is plausible.

The COBE Bulge has a scale height of $\sim$ 150 pc, and could be
interpreted as evidence for a separate component superimposed on top of
the optical Bulge. Further evidence that this may indeed be a separate
entity (the 'nucleus'?)  is provided by the fact that both luminosity
and kinematical models (eg Kent 1992) which describe well the outer
bulge under-predict the light in the very central part of the Galactic
Bulge. At the position of the SGR-I window, at 3 COBE scale lengths
from the Galactic Centre we are just beginning to pick up the IR
Bulge.  It is quite plausible that this centrally condensed structure
is considerably more metal-rich than the underlying, larger, optical
Bulge which is the main contributor to the stellar population(s) in
Baade's window, at $<[Fe/H]> = -0.3$ dex.

These results are supported by recent studies of OH/IR stars
(Sevenster, et al. 1997). OH/IR stars, which are oxygen-rich, cool giant
stars in their final stages of evolution, trace basically all stars
with a main-sequence mass of $1-6 M_{\sun}$ and can be reliably
detected through their OH maser emission at 1612 MHz. This makes them
ideal tracers of the underlying stellar population. They are
intermediate age or old, and should therefore be dynamically relaxed
and trace the global gravitational potential. These stars are
distributed in the central Bulge with a scale height $\la 100$ pc
(Sevenster et al. 1997).

We also note the fact that star-forming regions such as Sgr B exist
shows that star-formation is still going on in the Galactic Bulge on
scales of 50-100 pc. This is yet further evidence that we could expect
the central parts of the Galactic Bulge to be more metal-rich than the
outer edges of the Galactic Bulge. There remains perhaps no more than
a semantic distinction between continuing formation of the central
Disk and the central Bulge on such small scale lengths and recent
times.

In conclusion, both large scale modeling of the dynamics in the Bulge
and specific tracers show there to be structure in the stellar
populations at scales below those of the optical Bulge. The central
parts are a highly dissipated self-enriching part of the Galaxy which
is at least partially young.

\section{Conclusions}

We demonstrate how deep images with high spatial resolutions, only
available since the advent of HST, enable us to determine the
properties of the Galactic Bulge stellar population(s). In this we
have used three measures;

\begin{itemize}

\item number counts - to search for a young stellar population

\item histograms - to search for a metal-rich population

\item comparison of ridge lines - to derive metallicities and look
for internal differences, i.e. gradients in age and metallicity

\end{itemize}

\noindent
In particular we  find that

\begin{itemize}

\item  our results are consistent with no significant 
young stellar population in the Bulge, some 500pc, or 2-5 scale
lengths, from the centre, 
contrary to some previous studies;

\item the Galactic Bulge has a mean metallicity 
equal to that of the old disk;

\item there is  marginal evidence for a central metallicity gradient;

\item the colour-magnitude diagram of NGC6553 is complex, requiring
care in photometric analyses of this cluster.

\end{itemize}

\begin{acknowledgements}
The UK HST support group in Cambridge, Rachel Johnson and Nial Tanvir,
are thanked for numerous discussions on HST/WFPC2 and the best way to
extract stellar photometry.

To produce Fig.\ref{harris_histogram} we have use the online version
of the Harris (1996) catalogue.

SF acknowledges financial support from the Swedish Natural Research
Council under their postdoc program.
\end{acknowledgements}

\end{document}